\documentclass[twocolumn,secnumarabic,amssymb, nobibnotes, superscriptaddress, nofootinbib,aps,prd]{revtex4-2}
\usepackage[colorlinks,citecolor=blue,linkcolor=blue,anchorcolor=blue,filecolor=blue, urlcolor=blue]{hyperref}
\usepackage{natbib}

\usepackage{amsmath}
\usepackage{adjustbox}
\usepackage{graphicx}
\usepackage{booktabs}       
\usepackage{amsfonts}      
\usepackage{nicefrac}       
\usepackage{microtype}
\usepackage{cleveref}
\usepackage{xcolor}
\usepackage{multirow}
\usepackage{makecell}
\usepackage{array}
\usepackage{orcidlink}
\usepackage{float}
\usepackage{lineno}
\usepackage{slashed}
\usepackage{bm}
\definecolor{dora}{RGB}{255,150,0}

\usepackage[normalem]{ulem}
\usepackage{caption}
\DeclareCaptionJustification{fulljustify}{\leftskip=0pt \rightskip=0pt \parfillskip=0pt plus 1fil}
\usepackage{tikz}
\usepackage{tikz-feynman}
\usepackage{caption}
\usepackage{subcaption}

\begin{document}
\author{Adamu Issifu \orcidlink{0000-0002-2843-835X}} 
\email{ai@academico.ufpb.br}
\affiliation{CFisUC, Department of Physics, University of Coimbra, 3004-516 Coimbra, Portugal}
\affiliation{Departamento de F\'isica, Instituto Tecnol\'ogico de Aeron\'autica, DCTA, 12228-900, S\~ao Jos\'e dos Campos, SP, Brazil} 
\affiliation{Laborat\'orio de Computa\c c\~ao Cient\'ifica Avan\c cada e Modelamento (Lab-CCAM), Brazil}

\title{A self-consistent Higgs-portal framework for dark matter–admixed neutron stars: Collider-motivated benchmarks meet multimessenger constraints}

\begin{abstract}
We investigate dark matter (DM)-admixed neutron stars (NSs) within a self-consistent single-fluid relativistic mean-field framework by extending the Higgs-portal model with a massive $Z^\prime$ vector mediator. The resulting density-dependent repulsive interaction dynamically couples the baryonic matter (BM) and DM sectors, allowing the DM content to be characterized by the global particle fraction, $F_\chi=N_\chi/N_B$, with the local DM density determined self-consistently as $n_\chi=F_\chi n_B$, thereby eliminating the need for the externally prescribed DM Fermi momentum adopted in previous single-fluid models. Using the NL3$\omega\rho$, DD2, and FSU2R EOSs, we show that increasing $F_\chi$ systematically softens the nuclear equation of state (EOS), reduces the maximum NS mass by up to $\sim29\%$, increases stellar compactness, and modifies the thermodynamic response of dense matter. We further derive exact analytical expressions for the adiabatic speed of sound, its density derivative, and the adiabatic index, providing a rigorous benchmark for assessing the causality and thermodynamic stability of DM-admixed EOSs. At the microscopic level, the BM--DM interaction is shown to be dominated by the repulsive $Z^\prime$ vector channel. Our framework establishes a direct connection between collider-motivated WIMP models and the multimessenger phenomenology of NSs, with potential implications for future gravitational-wave and X-ray observations.
\end{abstract}

\maketitle

\section{Introduction}
The theoretical and experimental search for dark matter (DM) and its fundamental properties, including its mass, coupling strengths, and interactions with ordinary (baryonic) matter (BM), has become one of the central challenges in modern cosmology, particle physics, astronomy, and astrophysics. However, the interest in DM search is driven by the overwhelming evidence for DM and its dominant contribution to the matter content of the Universe. According to the standard cosmological model, the Universe consists of approximately $27\%$ DM, $68\%$ dark energy, and only $5\%$ BM. These estimates are supported by a wide range of observations, including galactic rotation curves~\cite{1980ApJ...238..471R,1981AJ.....86.1825B}, gravitational lensing~\cite{Clowe:2006eq}, galaxy cluster dynamics~\cite{Clowe:2006eq}, large-scale structure formation~\cite{Springel:2005nw,Planck:2015fie}, cosmic microwave background anisotropies~\cite{Planck:2015fie}, and observations of colliding galaxy clusters such as the Bullet Cluster. Despite the strong cosmological and astrophysical evidence for DM, its direct detection in terrestrial experiments has remained elusive, with no conclusive signals observed despite decades of dedicated searches and significant experimental investment~\cite{XENON:2018voc, XENON:2025vwd, LZ:2024zvo, PandaX:2024qfu}.

As a result, indirect DM searches have increasingly focused on investigating how DM capture, accumulation, and the possible formation of DM-admixed compact objects modify the structure and evolution of neutron stars (NSs) and white dwarfs~\cite{Kouvaris:2007ay, Issifu:2024htq, Lopes:2024ixl, deLavallaz:2010wp}. Owing to their extreme densities and high escape velocities, compact stars are expected to efficiently capture and retain DM over their lifetimes, making them promising laboratories for probing DM physics \cite{Bramante:2023djs}. In particular, compact stars residing in regions of high DM density, such as dwarf galaxies, galactic centers, and galaxy clusters, are expected to accumulate significantly more DM than their counterparts in low-density environments \cite{DelPopolo:2020hel}. Comparing the structural and observational properties of these populations therefore provides a unique opportunity to constrain the nature of DM and its interactions with BM \cite{Grippa:2024ach}. 

The uncertain nature of DM and its interactions with BM presents a major challenge for modeling its impact on compact stars. Consequently, two phenomenological frameworks are commonly employed. The first assumes that DM interacts with BM only gravitationally, or so weakly that the two components remain dynamically distinct, and describes the stellar structure using the two-fluid Tolman--Oppenheimer--Volkoff (TOV) formalism \cite{Shakeri:2022dwg, Issifu:2025jac, Thakur:2023aqm, Issifu:2025gsq, Kain:2021hpk, Biesdorf:2024dor, Rezaei:2016zje}. This approach is well suited to collisionless or weakly interacting DM scenarios, although the DM distribution is typically prescribed through model assumptions rather than derived from capture and thermalization processes. The second framework assumes sufficiently strong non-gravitational interactions between DM and BM, allowing the two components to thermalize and co-move as a single-fluid described by the standard TOV equations \cite{Bertoni:2013bsa, Bramante:2013nma}. In this picture, DM particles captured through repeated scattering lose kinetic energy, become gravitationally bound by the star's deep gravitational potential, and gradually settle toward the stellar core, where they accumulate over time. The resulting DM admixture modifies the microscopic composition, equation of state (EOS), and macroscopic structure of the star \cite{Issifu:2025qqw, Das:2020vng, Gresham:2018rqo, Sahoo:2025rqw, Das:2026hbu}.

Conventional studies of DM-admixed neutron stars (DANSs) within the single-fluid framework typically prescribe the DM Fermi momentum, $k_F^\chi$, as an external input parameter~\cite{Bramante:2023djs, Grippa:2024ach}. The most widely adopted particle-physics realization is the Higgs-portal model~\cite{Bhat:2019tnz, Lopes:2025jyz, Issifu:2025jac}, which successfully accounts for the observed cosmological DM relic abundance and provides a useful framework for investigating DM effects in compact stars. However, its implementation in NS studies remains largely phenomenological. In particular, treatments that prescribe a fixed $k_F^\chi$ effectively lead to the \emph{ad hoc} freezing of the local DM density, $n_\chi$,  and decouple it from the stellar environment \cite{Bhat:2019tnz}. Therefore, this assumption lacks a clear astrophysical motivation because it neglects the capture, thermalization, and gravitational settling of DM particles inside the star. 

In realistic capture scenarios, repeated DM--BM scattering causes DM particles to lose kinetic energy and become gravitationally bound, leading to a local DM density that follows the baryonic density profile rather than remaining externally prescribed. Consequently, the conventional fixed-$k_F^\chi$ prescription cannot consistently describe the density evolution of the DM component expected in NSs. To overcome this limitation, we extend the conventional Higgs-portal model by introducing a neutral massive vector mediator, $X_\mu$, associated with a hidden $U(1)_X$ gauge symmetry \cite{Langacker:2008yv}. Analogous to the $\omega$-meson in relativistic mean-field theory \cite{Menezes:2021jmw, Vretenar:2005zz, Oertel:2016bki}, $X_\mu$ generates a density-dependent repulsive vector mean-field that dynamically couples the BM and DM sectors by modifying their chemical potentials. This interaction enables the local DM density to evolve self-consistently according to $n_\chi=F_\chi n_B$, where $F_\chi=N_\chi/N_B$ is the global DM particle fraction, while maintaining the renormalizable structure of the underlying theory. The resulting framework provides a physically motivated description of DM accumulation in NSs that eliminates the need to prescribe the DM density or $k_F ^\chi$ externally \cite{Hajkarim:2024ecp, Issifu:2026diw}.

We constrain the particle-physics sector using benchmark values motivated by collider and direct-detection experiments~\cite{CMS:2021dzg, CMS:2020ulv, ATLAS:2024kpy, XENON:2018voc, XENON:2025vwd, LZ:2024zvo, PandaX:2024qfu}. The DM mass is chosen within the WIMP paradigm, the mediator mass is fixed to that of a heavy $Z^\prime$ boson, and the interaction couplings are constrained by current experimental limits, leaving the global $F_\chi$ as the only free astrophysical parameter describing the amount of DM accumulated inside an NS. Unlike the scalar Higgs portal single-fluid DANS phenomenology, which requires fixing $n_\chi$ independently of the stellar environment, we determine $n_\chi$ self-consistently through $n_\chi=F_\chi n_B$. Consequently, both $n_\chi$ and $k_F^\chi$ evolve naturally with $n_B$, consistent with the expected capture, thermalization, and gravitational settling of DM particles, while the BM sector remains in $\beta$-equilibrium and satisfies charge neutrality. The resulting EOS is then used to solve the TOV equations. The foundation of this framework was first proposed in \cite{Issifu:2026diw} to investigate mirror DM with particle masses set to be equal to the nucleon mass.

The paper is organized as follows. In \cref{theory}, we introduce the theoretical framework, presenting the DM model in \cref{dmm} and the BM model in \cref{bmm}. The EOS is developed in \cref{eos}, where \cref{equc} outlines the equilibrium conditions, \cref{energyd} presents the energy decomposition, \cref{cs2mod} derives the adiabatic speed of sound, \cref{sc2den} derives its density dependence, and \cref{adiab} presents the corresponding adiabatic index. The results, presented in \cref{result}, are divided into microphysics (\cref{mcphy}), macrophysics (\cref{macphy}), and DM--BM interaction properties (\cref{dmbmint}). Finally, our conclusions are given in \cref{conc}.

\section{Theoretical framework}\label{theory}

\begin{table}[t!]
\centering
\caption{RMF model parameters used in this work. Masses are given in MeV.}
\begin{tabular}{c c c}
\hline
Parameter & NL3$\omega\rho$~\cite{Providencia:2012rx} & FSU2R~\cite{Tolos:2016hhl,Tolos:2017lgv} \\
\hline
$m_N$      & 938.0   & 939.0 \\
$m_\sigma$ & 508.194 & 497.479 \\
$m_\omega$ & 782.501 & 782.500 \\
$m_\rho$   & 763.0   & 763.0 \\
$g_\sigma$ & 10.217  & 10.372 \\
$g_\omega$ & 12.868  & 13.505 \\
$g_\rho$   & 11.277  & 14.367 \\
$b$        & 0.00205 & 0.00165 \\
$c$        & -0.00265 & -0.00028 \\
$\xi$      & 0.0     & 0.024 \\
$\Lambda_\omega$ & 0.03 & 0.045 \\
$n_0$ (fm$^{-3}$) & 0.148 & 0.1505 \\
\hline
\end{tabular}
\label{tabnl3}
\end{table}

\subsection{Dark matter models}\label{dmm}
\begin{table}[t!]
\centering
\caption{Parameters of the density-dependent relativistic mean-field DD2 interaction adopted in this work from Ref.~\cite{Typel:2009sy}.}
\begin{tabular}{lccc}
\hline
Parameter & $\sigma$ & $\omega$ & $\rho$ \\
\hline
$m_i$ (MeV)                  & 546.212459 & 783.000000 & 763.000000 \\
$g_i(n_{\rm sat})$      & 10.686681  & 13.342362  & 3.626940 \\
$a_i$                        & 1.357630   & 1.369718   & 0.518903 \\
$b_i$                        & 0.634442   & 0.496475   & --- \\
$c_i$                        & 1.005358   & 0.817753   & --- \\
$d_i$                        & 0.575810   & 0.638452   & --- \\
\hline
\multicolumn{4}{c}{}\\[-1.5ex]
\multicolumn{2}{l}{$m_N$ (MeV)}              & \multicolumn{2}{c}{939} \\
\multicolumn{2}{l}{$n_{0}$ (fm$^{-3}$)} & \multicolumn{2}{c}{0.1491} \\
\hline
\end{tabular}
\label{tab:DD2}
\end{table}

The Higgs-portal Lagrangian is given by
\begin{align}
\mathcal{L}_{\rm HP} &=\bar{\chi}\left(i\gamma^\mu\partial_\mu-\left(m_\chi-g_h h\right)\right)\chi\nonumber\\ &+\frac{1}{2}\left(\partial^\mu h\,\partial_\mu h-m_h^2h^2\right)
+\sum_N\frac{f\,m_N}{v}\bar{\psi}_Nh\psi_N ,
\label{HP}
\end{align}
where $\chi$ denotes the dark fermion, $h$ is the Higgs field, $m_\chi$ and $m_h = 125\, \rm GeV$ are the DM and Higgs masses, respectively, $g_h$ is the Higgs--DM Yukawa coupling, $m_N$ is the nucleon mass, $v=246\,\mathrm{GeV}$ is the Higgs vacuum expectation value, and $f$ parameterizes the effective Higgs--nucleon form factor. We consider a WIMP, commonly identified with the neutralino in supersymmetric extensions of the Standard Model, and adopt a benchmark mass of $m_\chi = 200\,\mathrm{GeV}$. For the Higgs--nucleon interaction, we use the scalar nucleon form factor $f=0.35$, consistent with determinations from lattice QCD~\cite{Czarnecki:2010gb}, the MILC Collaboration~\cite{Toussaint:2009pz}, and the ATLAS Collaboration~\cite{ATLAS:2015ciy}. The Higgs--DM Yukawa coupling is fixed to $g_h=0.07$, which lies well within the phenomenologically viable range $0.001 \leq g_h \leq 0.1$~\cite{Panotopoulos:2017idn}. These benchmark parameters yield a spin-independent DM--nucleon scattering cross-section of the order of $\sigma_{\chi N}^{\rm SI}\sim10^{-47}\,\mathrm{cm}^2$, consistent with the current upper limits reported by direct-detection experiments, including XENON1T/XENONnT, LZ, and PandaX-4T~\cite{XENON:2018voc,XENON:2025vwd,LZ:2024zvo,PandaX:2024qfu}.

To obtain a self-consistent treatment of $n_\chi$, we extend the Higgs-portal Lagrangian, $\mathcal{L}_{\rm HP}$, by introducing a neutral massive gauge boson, $X_\mu$, associated with a hidden $U(1)_X$ symmetry. In this work, $X_\mu$ is identified with a dark $Z'$ boson, a well-motivated vector mediator extensively investigated in both theoretical studies and collider searches, with TeV-scale benchmark masses typically in the range $1$--$5\,\mathrm{TeV}$ \cite{Langacker:2008yv, CMS:2021dzg, CMS:2020ulv, ATLAS:2024kpy}. The inclusion of this mediator generates an effective interaction between the dark fermions and the visible baryons through vector-current couplings, providing a mechanism for the simultaneous evolution of the dark and baryonic sector densities. The Lagrangian density for the neutral massive gauge boson is given by: 
\begin{align}
\mathcal{L}_{X}&=-\frac14 X_{\mu\nu}X^{\mu\nu}+\frac12 m_X^2X_\mu X^\mu-\sum_N g_{NX}\bar{\psi}_N\gamma^\mu\psi_NX_\mu\nonumber\\
&-g_{\chi X}\bar{\chi}\gamma^\mu\chi X_\mu ,
\label{Vector}
\end{align}
where
\begin{align}
X_{\mu\nu}=\partial_\mu X_\nu-\partial_\nu X_\mu
\end{align}
is the field-strength tensor of the vector mediator, $m_X$ is its mass, while $g_{NX}$ and $g_{\chi X}$ denote the vector couplings to the baryonic and dark sectors, respectively. Physically, $X_\mu$ plays a role analogous to the $\omega$-meson in RMF theory, generating a vector mean-field that modifies the chemical potentials of both sectors and enables a fully self-consistent treatment of the BM--DM system. Within the mean-field approximation, $X_\mu\rightarrow(X_0,0,0,0)$. Similar theories have been introduced in \cite{Hardy:2022ufh, Capanelli:2024pzd} to study lighter vector particles like dark photons.  

%CMS:2021dzg, CMS:2020ulv, ATLAS:2024kpy,
Following the benchmark spin-1 vector mediator scenario adopted by the ATLAS Collaboration for simplified DM models at the $95\%$ CR~\cite{ATLAS:2024kpy,ATLAS:2021ovy}, we fix the quark--mediator and DM--mediator couplings to the recommended benchmark values $g_{qX}=0.25$ and $g_{\chi X}=1$, respectively. Assuming universal quark couplings, $g_{qX}=g_u=g_d$, the effective nucleon--mediator coupling is obtained in the additive quark model by coherently summing the vector current over the three valence quarks, yielding the approximation $g_{NX}\simeq 3g_{qX}=0.75$. Since collider searches probe mediator masses over the TeV range rather than a single benchmark mass, we adopt $m_X=1.5~\mathrm{TeV}$ as a representative heavy spin-1 mediator consistent with the parameter space explored by ATLAS and with generic $Z'$ scenarios discussed in \cite{Langacker:2008yv}. With the particle-physics sector fixed by collider-motivated benchmarks, the global DM fraction, $F_\chi$, remains the only free astrophysical parameter, encoding the uncertain amount of DM accumulated inside the NS.

\subsection{Baryonic Matter}\label{bmm}

In the BM sector, we employ three relativistic mean-field (RMF) EOS that are widely used in the literature: the third version of the nonlinear RMF model with an $\omega$--$\rho$ coupling, NL3$\omega\rho$~\cite{Providencia:2012rx,Horowitz:2000xj}; the Florida State University model with reduced NS radii, FSU2R~\cite{Tolos:2017lgv}; and the density-dependent RMF model, version~2 (DD2)~\cite{Typel:2009sy}. The coupling constants for the NL3$\omega\rho$ and FSU2R parameterizations are listed in \cref{tabnl3}, while those of DD2 are given in \cref{tab:DD2}. The corresponding BM Lagrangian density is presented in \cref{lgbm}. 

These three models were selected because they span a broad range of microscopic assumptions and macroscopic NS properties, allowing us to assess the model dependence of the EOS in the presence of DM. In particular, NL3$\omega\rho$ is a nonlinear RMF model with a relatively stiff EOS that predicts larger NS radii and higher maximum masses. In contrast, FSU2R incorporates non-negligible stronger isoscalar--isovector interactions, producing a softer EOS and consequently smaller stellar radii while remaining compatible with the observed $2\,M_\odot$ constraint. The DD2 model employs density-dependent meson--nucleon couplings instead of nonlinear meson self-interactions and exhibits an intermediate stiffness, providing a useful comparison between nonlinear and density-dependent RMF descriptions.

\begin{widetext}
     The BM Lagrangian density:
\begin{align} \label{lgbm}
\mathcal{L}_B &=  \bar{\psi}_N \left[i\gamma^\mu\partial_\mu-\gamma^0 \left(  g_\omega \omega_0 + g_\rho I_{3\rho} \rho_{03} \right)- \left(m_N - g_\sigma \sigma_0 \right)\right] \psi_N - \frac{1}{2} m_\sigma^2 \sigma^2_0 + \frac{1}{2} m_\omega^2 \omega_0^2 + \frac{1}{2} m_\rho^2 \rho_{03}^2 \nonumber\\&
- \frac{1}{3} b\, m_N \, g_\sigma^3 \sigma^3_0- \frac{1}{4} c\, g_\sigma^4 \sigma^4_0+ \frac{\xi}{4!} g_\omega^4\omega_0^4+ \Lambda_\omega g_\rho^2 g_\omega^2 \rho_{03}^2\omega_0^2, \nonumber\\
\mathcal{L}_{l} &=\bar{\psi}_{l}\left(i\gamma^\mu\partial_\mu -m_{l}\right)\psi_{l}.
\end{align}
\end{widetext}

The field $\psi_N$ ($\bar{\psi}_N$) denotes the Dirac spinor for the nucleon doublet (protons and neutrons) with bare nucleon mass $m_N$, while $\gamma^\mu$ are the Dirac matrices and $I_{3\rho}=\pm 1/2$ is the third component of the isospin. The coupling constants $g_\sigma$, $g_\omega$, and $g_\rho$ describe the interactions of the nucleons with the scalar $\sigma$, vector $\omega$, and isovector-vector $\rho$ mesons, whose masses are denoted by $m_\sigma$, $m_\omega$, and $m_\rho$, respectively. In nonlinear relativistic mean-field models, the parameters $b$, $c$, $\xi$, and $\Lambda_\omega$, together with the meson--nucleon couplings, determine the bulk properties of nuclear matter. Specifically, the nonlinear scalar self-interaction parameters $b$ and $c$ primarily control the incompressibility near saturation density, the $\omega$-meson self-interaction parameter $\xi$ softens the EOS at high densities, and the mixed $\omega$--$\rho$ coupling $\Lambda_\omega$ regulates the density dependence of the symmetry energy, particularly its slope at saturation \cite{Providencia:2012rx, Tolos:2016hhl, Tolos:2017lgv}. $\psi_l$ ($\bar{\psi}_l$) are the free lepton fields with mass $m_l$ that are introduced in the stellar matter to ensure charge neutrality and $\beta$-equilibrium.

For the DD2 model, the meson--nucleon couplings $g_\sigma$, $g_\omega$, and $g_\rho$ depend explicitly on the baryon density according to
\begin{equation}
g_i(n_B)=g_i(n_0)\,h_i(x), \qquad
x=\frac{n_B}{n_0}, \qquad
i=\sigma,\omega,\rho,
\label{eq:dd_coupling}
\end{equation}
where $g_i(n_0)$ denotes the coupling constant at the saturation density $n_0$. The density dependence of the $\sigma$- and $\omega$-meson couplings is parameterized as
\begin{equation}
h_i(x)=a_i\,
\frac{1+b_i(x+d_i)^2}
{1+c_i(x+d_i)^2},
\qquad i=\sigma,\omega,
\label{eq:ddh_h}
\end{equation}
while the $\rho$-meson coupling is given by
\begin{equation}
h_\rho(x)=\exp\!\left[-a_\rho(x-1)\right].
\label{eq:hy}
\end{equation}
Unlike nonlinear relativistic mean-field models, the DD2 parameterization does not include nonlinear meson self-interactions or mixed meson interaction terms, so that the corresponding coefficients $b$, $c$, $\xi$, and $\Lambda_\omega$ vanish \cite{Typel:2009sy}. Also, the effective chemical potential and the associated pressure derived from this model are modified by the rearrangement self-energy term $\Sigma_R$ to ensure thermodynamic self-consistency:
\begin{equation}
\Sigma_R^B=\frac{\partial g_{\omega}}{\partial n_B}\omega_0\,n_B
+\frac{\partial g_{\rho}}{\partial n_B}\rho_{03}\,n_{3B}- \frac{\partial g_{\sigma}}{\partial n_B}\sigma\,n_{s}^B,
\end{equation}
where $n_{sB}$ is the scalar density.

\section{The equation of state}\label{eos}
This section presents the derivation of the EOS for DANSs, which forms the theoretical foundation of this work. We focus on the EOS of the DM sector and its coupling to the BM sector. Since the EOS of the BM sector has been extensively studied in the literature, we do not repeat those derivations here and instead refer the interested reader to Refs.~\cite{Menezes:2021jmw, Vretenar:2005zz, Oertel:2016bki}. The complete Lagrangian density for the study becomes:
\begin{equation}\label{netl}
    \mathcal{L} = \mathcal{L}_B +\mathcal{L}_X +\mathcal{L}_{HP} +\mathcal{L}_l.
\end{equation}
Applying the Euler--Lagrange equation to variation in $\bar{\psi}_N$ yields,
\begin{align}
\Big[&i\gamma^\mu\partial_\mu-\gamma^0\left(g_\omega\omega_0+g_\rho I_{3\rho}\rho_{03}+g_{NX}X_0\right)
\nonumber\\
&-\left(m_N-g_\sigma\sigma_0-\frac{fm_N}{v}h_0\right)\Big]\psi_N=0,
\label{DiracB}
\end{align}
where $h_0$, $X_0$ are the mean-field versions of $h$ and $X_\mu$. Similarly, variation in $\bar{\chi}$, satisfies
\begin{align}
\left[i\gamma^\mu\partial_\mu-g_{\chi X}\gamma^0X_0-\left(m_\chi-g_hh_0\right)\right]\chi=0.
\label{DiracDM}
\end{align}
The equation of motion of the Higgs field gives
\begin{align}
m_h^2h_0=g_h\langle\bar{\chi}\chi\rangle+\frac{fm_N}{v}\langle\bar{\psi}_N\psi_N\rangle.
\end{align}
Introducing the scalar densities,
\begin{equation}
    n_{s}^{\chi}=\langle\bar{\chi}\chi\rangle=
    \frac{\gamma_{\chi}}{2\pi^{2}}
    \int_{0}^{k_{F}^{\chi}}
    \frac{m_{\chi}^{*}}
    {\sqrt{k^{2}+m_{\chi}^{*2}}}\,
    k^{2}\,dk,
\end{equation}
where $n_{s}^{\chi}$ is the DM scalar density, $\gamma_{\chi}=2$ is the spin degeneracy factor of the dark fermion, and
\begin{equation}
    k_{F}^{\chi}=\left({3\pi^{2}n_{\chi}}\right)^{1/3} = \left({3\pi^{2}F_\chi n_{B}}\right)^{1/3}
\end{equation}
is the dark Fermi momentum, with $n_\chi = F_\chi n_B$ denoting the dark fermion number density. The scalar baryon density is also given by
\begin{align}
n_s^B=\langle\bar{\psi}_N\psi_N\rangle.
\end{align}
The Higgs field equation becomes
\begin{align}
m_h^2h_0=g_hn_s^\chi+\frac{fm_N}{v}n_s^B.\label{HiggsMF}
\end{align}
Variation with respect to the vector field in \eqref{Vector} gives
\begin{align}
m_X^2X_0=g_{NX}n_B+g_{\chi X}n_\chi,
\label{Xfield}
\end{align}
\begin{align}
n_B=\langle\psi_N^\dagger\psi_N\rangle,\qquad n_\chi=\langle\chi^\dagger\chi\rangle.
\end{align}
The $\sigma$, $\omega$, and $\rho$ field equations retain their usual RMF form,
\begin{align}
m_\sigma^2\sigma_0+bm_Ng_\sigma^3\sigma_0^2+cg_\sigma^4\sigma_0^3=g_\sigma n_s^B,
\end{align}
\begin{align}
m_\omega^2\omega_0+\frac{\xi}{6}g_\omega^4\omega_0^3+2\Lambda_\omega g_\omega^2g_\rho^2\omega_0\rho_{03}^2=g_\omega n_B,
\end{align}
and
\begin{align}m_\rho^2\rho_{03}+2\Lambda_\omega g_\omega^2g_\rho^2\omega_0^2 \rho_{03}=g_\rho n_3, 
\end{align}

where 
\begin{equation}
    n_3 = \sum_{i\in n,p}I_{3\rho}\dfrac{\gamma_i}{6\pi^2}k_{F_i}^3.
\end{equation}
The effective baryon and DM masses are
\begin{align}
m_N^*&=m_N-g_\sigma\sigma_0-\frac{fm_N}{v}h_0,\label{ncefm}
\\
m_\chi^*&=m_\chi-g_hh_0.
\label{EffMass}
\end{align}
The proton, neutron, and DM chemical potentials are given by
\begin{align}
\mu_{p,n}&=\sqrt{k_{F_{p,n}}^2+m_N^{*2}}+g_\omega\omega_0+I_{3\rho}g_\rho\rho_{03}+g_{NX}X_0,
\\
\mu_\chi&=\sqrt{k_{F\chi}^2+m_\chi^{*2}}+g_{\chi X}X_0
\label{ChemPot}
\end{align}
where $\mu^*_{p,n} =\sqrt{k_{F_{p,n}}^2+m_N^{*2}}$ and $\mu_\chi^* = \sqrt{k_{F\chi}^2+m_\chi^{*2}}$ are the effective chemical potentials of the BM and the DM sectors, respectively. 

The shift in the chemical potentials induced by the vector mediator $X_\mu$ represents a density-dependent repulsive mean-field potential that dynamically couples the BM and DM sectors. As shown in \cref{Xfield}, the $X_0$ is determined self-consistently by the baryonic and DM densities, shifting the nucleon and DM chemical potentials by $g_{NX}X_0$ and $g_{\chi X}X_0$, respectively. Unlike the scalar Higgs portal, this vector interaction increases the energy cost of compressing BM and DM into the same volume, thereby regulating their coexistence in dense matter. Although the common baryonic shift cancels explicitly from the $\beta$-equilibrium condition, its density dependence modifies the coupled field equations, altering the effective masses, Fermi momenta, meson fields, and charge-neutrality condition. Consequently, the BM and DM sectors remain coupled through a nonlinear feedback mechanism, allowing the DM density, EOS, and NS structure to evolve self-consistently rather than being prescribed by an \emph{ad hoc} DM distribution. A related single-fluid model was considered in Ref.~\cite{Kumar:2026hoq}; however, the DM density was prescribed through a fixed $k_F^\chi$, effectively treating the local DM Fermi momentum as an external parameter. A variant in which the vector mediator couples exclusively to the DM sector was studied in Ref.~\cite{Sotani:2025lzy} in a two-fluid scenario.

The EOS of DANSs is obtained from \cref{netl}, given rise to \cref{engd,pres}
\begin{widetext}
\begin{align}
\varepsilon_{\rm tot} &=\sum_{i=n,p}\frac{\gamma_B}{2\pi^2}\int_0^{k_{Fi}} dk\,k^2\sqrt{k^2+m_N^{*2}}+\sum_{l=e,\mu}\frac{1}{\pi^2}\int_0^{k_{Fl}} dk\,k^2\sqrt{k^2+m_l^2}
\nonumber\\
&+\frac{\gamma_\chi}{2\pi^2}\int_0^{k_{F\chi}} dk\,k^2\sqrt{k^2+m_\chi^{*2}}+\frac12m_\sigma^2\sigma_0^2+\frac13b(g_\sigma\sigma_0)^3+\frac14c(g_\sigma\sigma_0)^4 +\frac12m_\omega^2\omega_0^2
\nonumber\\
&+\frac{\xi}{8}g_\omega^4\omega_0^4+\frac12m_\rho^2\rho_{03}^2+3\Lambda_\omega g_\omega^2g_\rho^2\omega_0^2\rho_{03}^2 +\frac12m_h^2h^2+\frac12m_X^2X_0^2, \label{engd}
\\
P_{\rm tot}&=\sum_{i=n,p}\frac{\gamma_B}{6\pi^2}\int_0^{k_{Fi}} dk\,\frac{k^4}{\sqrt{k^2+m_N^{*2}}}+\sum_{l=e,\mu}\frac{1}{3\pi^2}\int_0^{k_{Fl}} dk\,\frac{k^4}{\sqrt{k^2+m_l^2}}
\nonumber\\
&+\frac{\gamma_\chi}{6\pi^2}\int_0^{k_{F\chi}} dk\frac{k^4}{\sqrt{k^2+m_\chi^{*2}}}-\frac12m_\sigma^2\sigma_0^2-\frac13b(g_\sigma\sigma_0)^3-\frac14c(g_\sigma\sigma_0)^4 +\frac12m_\omega^2\omega_0^2+\frac{\xi}{24}g_\omega^4\omega_0^4
\nonumber\\
&+\frac12m_\rho^2\rho_{03}^2+\Lambda_\omega g_\omega^2g_\rho^2\omega_0^2\rho_{03}^2-\frac12m_h^2h^2 +\frac12m_X^2X_0^2, \label{pres}
\end{align}
\end{widetext}
where $\varepsilon_{\rm tot}$ and $P_{\rm tot}$ denote the total energy density and pressure of the coupled BM--DM system, respectively. The quantity $\gamma_B=2$ is the nucleon spin degeneracy factor, while $k_{Fl}$ is the lepton Fermi momentum. Throughout this work, the leptonic sector consists of electrons ($e$) and muons ($\mu$).

\subsection{Equilibrium conditions and particle number conservation}\label{equc}
The stellar matter satisfies $\beta$-equilibrium, $\mu_n=\mu_p+\mu_e$, and charge neutrality, $n_p=n_e+n_\mu$, in the BM sector. To characterize the DM content, we impose a global particle fraction,
\begin{equation}
    F_\chi=\frac{N_\chi}{N_B},
\end{equation}
where the total number of baryons and DM particles is
\begin{align}
N_\chi &=\int_0^R4\pi r^2\left(1-\frac{2M(r)}{r}\right)^{-1/2}n_\chi(r)\,dr,
\\
N_B &=\int_0^R4\pi r^2\left(1-\frac{2M(r)}{r}\right)^{-1/2}n_B(r)\,dr,
\end{align}
where the proper volume element is
\begin{equation}
dV_p=4\pi r^2\left(1-\frac{2M(r)}{r}\right)^{-1/2}dr.
\end{equation}
Within the present single-fluid framework, we assume that the captured DM has fully thermalized with the BM and co-moves with it, so both components occupy the same infinitesimal proper fluid element. Hence,
\begin{equation}
dN_\chi=n_\chi\,dV_p,\qquad dN_B=n_B\,dV_p.
\end{equation}
Furthermore, because the two species are assumed to remain perfectly mixed after thermalization, the local dark-to-baryon particle ratio is conserved in every co-moving fluid element,
\begin{equation}
\frac{dN_\chi}{dN_B}=F_\chi,
\end{equation}
which gives
\begin{equation}\label{fd}
n_\chi=F_\chi n_B.
\end{equation}
$F_\chi$ provides an astrophysically meaningful measure of the total DM accumulated by the NS through long-term capture from its galactic environment \cite{deLavallaz:2010wp}. This relation is imposed directly in the EOS calculation, allowing the local $n_\chi$ and $k_F^\chi$ to be determined self-consistently from $n_B$ rather than prescribing a fixed $k_F^\chi$. Consequently, the DM distribution follows the $n_B$ profile and is naturally concentrated toward the stellar core, where the gravitational potential is deepest.

The total rest mass, $M_0$, of the star, accounting for both the BM and DM components, can be expressed as
\begin{equation}\label{mrest}
M_0=N_Bm_N+N_\chi m_\chi
    =N_Bm_N\left[1+F_\chi\left(\frac{m_\chi}{m_N}\right)\right].
\end{equation}

\subsection{Self-consistent BM--DM interaction energy}\label{energyd}
The interaction between the baryonic and dark sectors consists of two contributions: the scalar interaction mediated by the Higgs portal and the vector interaction mediated by the massive dark boson $X_\mu$. The total DM--BM interaction is obtained from the sum of these two contributions. From \cref{engd} the vector contribution to the energy density is given by:
\begin{equation}
    \varepsilon_X =\dfrac{1}{2}m_X^2X_0^2.
\end{equation}
From the $X_0$ equation of motion \cref{Xfield} we can subtitute $X_0$ and write:
\begin{align}\label{vself}
    \varepsilon_X &= \dfrac{\left(g_{NX}n_B + g_{\chi X}n_\chi\right)^2}{2m_X^2}\nonumber\\
    &= \dfrac{g_{NX}^2n_B^2}{2m_X^2} + \dfrac{g_{\chi X}^2n_\chi^2}{2m_X^2}+\dfrac{g_{NX}g_{\chi X}n_Bn_\chi}{m_X^2}.
\end{align}
Therefore, the vector interaction density between the two sectors becomes 
\begin{equation}\label{vecden}
    \varepsilon_{\rm int}^{(X)} = \dfrac{g_{NX}g_{\chi X}n_Bn_\chi}{m_X^2}.
\end{equation}
The scalar contribution from \cref{engd} gives
\begin{equation}
    \varepsilon_h = \dfrac{1}{2}m_h^2h_0^2,
\end{equation}
substituting \cref{HiggsMF} gives,
\begin{align}\label{scself}
    \varepsilon_h &= \dfrac{\left(\frac{fm_N}{v}n_s^B+g_hn_s^\chi\right)^2}{2m_h^2}\nonumber\\
    &= \dfrac{\frac{f^2m_N^2}{v^2}}{2m_h^2}(n_s^B)^2 + \dfrac{g_h^2(n_s^\chi)^2}{2m_h^2} + \dfrac{fm_Ng_h}{vm_h^2}n_s^Bn_s^\chi. 
\end{align}
Thus, the scalar DM--BM interaction becomes
\begin{equation}\label{sclden}
    \varepsilon_{\rm int}^{(h)}= \dfrac{fm_Ng_h}{vm_h^2}n_s^Bn_s^\chi.
\end{equation}
The first and second terms in the last lines of \cref{vself,scself} correspond to the baryonic and DM self-energy densities, respectively, while the cross terms quantify the interaction energy density between the BM and DM sectors. This decomposition allows the scalar and vector-mediated contributions to the BM--DM interaction to be identified and analyzed separately. The net interaction of the DM--BM becomes,
\begin{equation}
    \varepsilon_{\rm int} = \varepsilon_{\rm int}^{(X)}+ \varepsilon_{\rm int}^{(h)}.
\end{equation}
The scalar Higgs interaction couples the baryonic and dark scalar densities, whereas the $Z^\prime$-mediated interaction couples their vector densities. The total interaction energy density therefore reflects the interplay between the scalar and vector contributions, whose relative strengths determine the net BM--DM interaction and its impact on the EOS.

While previous studies of scalar- and vector-mediated DM admixture in NSs have qualitatively shown the dominance of the vector interaction~\cite{Guha:2021njn, Kumar:2026hoq}, the present framework enables a quantitative comparison by deriving explicit expressions for the scalar and vector interaction energy densities, $\varepsilon_{\rm int}^{(h)}$ and $\varepsilon_{\rm int}^{(X)}$ (see \cref{vecden,sclden}) from the mean-field solutions. The decomposition shows that the vector contribution scales predominantly with $n_Bn_\chi$ and is largely insensitive to the nuclear EOS, since $n_\chi = F_\chi n_B$, whereas the scalar contribution depends on the EOS-dependent scalar densities, $n_s^B$ and $n_s^\chi$ (see \cref{int}). It therefore provides a microscopic interpretation of the EOS softening by quantifying the relative contributions of the two interaction channels throughout the star and establishes a framework for comparing different DM portal models.

\subsection{Speed of sound modification in self-consistent DANSs}\label{cs2mod}

The total pressure and energy density are
\begin{align}\label{topend}
P_{\rm tot}&=P_B+P_\chi + P_X +P_h,\\
\varepsilon_{\rm tot}&=\varepsilon_B+\varepsilon_\chi +\varepsilon_X + \varepsilon_h,
\end{align}
from which the adiabatic speed of sound follows as
\begin{equation}
c_s^2=\frac{dP_{\rm tot}}{d\varepsilon_{\rm tot}}=\frac{dP_B+dP_\chi + dP_X+ dP_h}{d\varepsilon_B+d\varepsilon_\chi + d\varepsilon_X +d\varepsilon_h}.
\label{cs2}
\end{equation}
In conventional single-fluid Higgs-portal models, $k_F^\chi$ is prescribed externally, $k_F^\chi={\rm const.}$, implying
\begin{equation}\label{cs2d}
\frac{dP_\chi}{dn_B}\simeq0,\qquad
\frac{d\varepsilon_\chi}{dn_B}\simeq0, \quad \frac{d\varepsilon_X}{dn_B}\simeq0 \quad \frac{d\varepsilon_h}{dn_B}\simeq0
\end{equation}
and therefore
\begin{equation}
c_s^2\simeq
\frac{dP_B/dn_B}
{d\varepsilon_B/dn_B},
\label{cs2a}
\end{equation}
so that the thermodynamic response is governed almost entirely by the BM sector. In the present framework, however, $n_\chi=F_\chi n_B$
and therefore
\begin{equation}
\frac{dn_\chi}{dn_B}=F_\chi.
\end{equation}
Consequently,
\begin{equation}
\frac{dP_D}{dn_B}=\frac{dP_D}{dn_\chi}\frac{dn_\chi}{dn_B},\qquad\frac{d\varepsilon_D}{dn_B}=\frac{d\varepsilon_D}{dn_\chi}\frac{dn_\chi}{dn_B},
\end{equation}
such that the DM sector contributes directly to the thermodynamic derivatives entering \cref{cs2}. $P_D=P_\chi + P_X + P_h$ and $\varepsilon_D=\varepsilon_\chi + \varepsilon_X + \varepsilon_h$ are the collective contributions for the DM sector pressure and energy densities, respectively.

The Higgs and vector mean-fields satisfy
\begin{equation}
h_0=h_0(n_s^B,n_s^\chi),
\qquad
X_0=X_0(n_B,n_\chi),
\end{equation}
while the RMF fields are denoted collectively by
\begin{equation}
\mathcal{M}\equiv(\sigma,\omega_0,\rho_{03}).
\end{equation}
Hence,
\begin{align}
P_B&=P_B(n_B,\mathcal{M},h_0,X_0),\\
P_D&=P_D(n_\chi,h_0,X_0),
\end{align}
giving
\begin{align}
\frac{dP_{\rm tot}}{dn_B}
&=\frac{\partial P_B}{\partial n_B}+\frac{\partial P_B}{\partial\mathcal{M}}\frac{d\mathcal{M}}{dn_B}+\frac{\partial P_B}{\partial h_0}\frac{dh_0}{dn_B}+\frac{\partial P_B}{\partial X_0}\frac{dX_0}{dn_B}
\nonumber\\
&
+\frac{\partial P_D}{\partial n_\chi}\frac{dn_\chi}{dn_B}+\frac{\partial P_D}{\partial h_0}\frac{dh_0}{dn_B}+\frac{\partial P_D}{\partial X_0}\frac{dX_0}{dn_B},
\end{align}
with analogous expressions for
$d\varepsilon_{\rm tot}/dn_B$.
The additional dependence on $h_0$, $X_0$, and $n_\chi$
distinguishes the present formulation from the conventional fixed-$k_F^\chi$ approach and provides the microscopic origin of the modified sound-speed profile.

\subsection{Density evolution of the adiabatic speed of sound}\label{sc2den}

For $\beta$-equilibrated matter,
\begin{equation}
P_{\rm tot}=P_{\rm tot}(n_B),
\qquad
\varepsilon_{\rm tot}=\varepsilon_{\rm tot}(n_B),
\end{equation}
so that
\begin{equation}
\frac{dc_s^2}{dn_B}=\frac{P_{\rm tot}''\varepsilon_{\rm tot}'-P_{\rm tot}'\varepsilon_{\rm tot}''}{\left(\varepsilon_{\rm tot}'\right)^2},
\label{dcs}
\end{equation}
which is valid for any EOS. From \cref{topend}, both the DM and BM species contribute to $P_{\rm tot}$ and $\varepsilon_{\rm tot}$ irrespective of the model parameterization.
% \begin{align}
% P_{\rm tot}&=P_B+P_\chi+P_h+P_X,\\
% \varepsilon_{\rm tot}
% &=
% \varepsilon_B+\varepsilon_\chi+\varepsilon_h+\varepsilon_X,
% \end{align}
%leading directly to \cref{dcs}. 
Therefore, the essential difference from fixed $k_F^\chi$ models and the present framework is that $n_\chi=F_\chi n_B$
leads to
\begin{equation}
\frac{dn_\chi}{dn_B}=F_\chi,
\qquad
\frac{d^2n_\chi}{dn_B^2}=0.
\end{equation}
Consequently,
\begin{equation}
k_F^\chi
=
(3\pi^2F_\chi n_B)^{1/3},
\end{equation}
satisfies
\begin{equation}\label{dkfx}
\frac{dk_F^\chi}{dn_B}=\frac{k_F^\chi}{3n_B},
\qquad
\frac{d^2k_F^\chi}{dn_B^2}=-\frac{2k_F^\chi}{9n_B^2},
\end{equation}
so that the DM kinetic, scalar, and vector sectors contribute explicitly to $P_{\rm tot}''$ and $\varepsilon_{\rm tot}''$. In contrast, a fixed $k_F^\chi$ gives
\begin{equation}
\frac{dk_F^\chi}{dn_B}=\frac{d^2k_F^\chi}{dn_B^2}=0,
\end{equation}
eliminating these contributions. \Cref{dcs} therefore identifies the microscopic origin of the modified density dependence of $c_s^2$ in the present self-consistent formulation.

\subsection{Adiabatic index}\label{adiab}

The adiabatic index,
\begin{equation}
\Gamma=
\frac{\varepsilon_{\rm tot}+P_{\rm tot}}{P_{\rm tot}}
\left(\frac{\partial P_{\rm tot}}{\partial\varepsilon_{\rm tot}}\right)_S
=
\frac{\varepsilon_{\rm tot}+P_{\rm tot}}{P_{\rm tot}}c_s^2,
\label{Gamma}
\end{equation}
quantifies the stiffness of dense matter against adiabatic compression and plays a central role in the stability and oscillation properties of compact stars \cite{Moustakidis:2016ndw,Koliogiannis:2018hoh,Glendenning2000}. Since the present framework modifies both the thermodynamic derivatives and the EOS, it also modifies $\Gamma$.

Single-fluid models with a fixed $k_F^\chi$ imply
\begin{equation*}
\frac{dk_F^\chi}{dn_B}=0,
\qquad
\frac{dn_\chi}{dn_B}=0,
\end{equation*}
so that the DM sector does not contribute to the thermodynamic derivatives (see \cref{cs2d,cs2a}). As a result,
\begin{equation}
c_{s,\rm fixed}^{2}=\frac{P_B'}{\varepsilon_B'},
\end{equation}
and
\begin{equation}
\Gamma_{\rm fixed}=\frac{\varepsilon_{\rm tot}+P_{\rm tot}}{P_{\rm tot}}\frac{P_B'}{\varepsilon_B'}.
\label{Gamma_fixed}
\end{equation}
Here, the DM contribution enters only through the algebraic prefactor $(\varepsilon_{\rm tot}+P_{\rm tot})/P_{\rm tot}$.

In the present framework, however, $dk_F^\chi/dn_B=k_F^\chi/3n_B$ (\cref{dkfx}), allowing the DM kinetic, scalar, vector, and interaction contributions to enter the thermodynamic derivatives. The corresponding $c_s^2$ and $\Gamma$ become
\begin{equation}
c_{s,\rm self}^{2}=\frac{P_B'+P_\chi'+P_X'+P_h'}{\varepsilon_B'+\varepsilon_\chi'+\varepsilon_X'+\varepsilon_h'},
\end{equation}
and
\begin{equation}
\Gamma_{\rm self}
=
\frac{\varepsilon_{\rm tot}+P_{\rm tot}}{P_{\rm tot}}
\frac{P_B'+P_\chi'+P_X'+P_h'}
{\varepsilon_B'+\varepsilon_\chi'+\varepsilon_X'+\varepsilon_h'}.
\label{Gamma_self}
\end{equation}

% Unlike the fixed-$k_F^\chi$ prescription, where the DM sector merely rescales $\Gamma$, the present formulation modifies both its magnitude and density dependence through $n_\chi$, $k_F^\chi$, $h_0$, and $X_0$. The $\Gamma$ therefore becomes an intrinsic property of the coupled BM--DM system rather than of the BM EOS alone. 
Comparing Eqs.~(\ref{Gamma_fixed}) and (\ref{Gamma_self}) shows that the current formulation allows the DM sector to participate directly in the thermodynamic response through the density dependence of $k_F^\chi$, the Higgs field, and the $Z^\prime$ vector mean-field. Therefore, $\Gamma$ becomes an intrinsic property of the coupled BM--DM system rather than being determined solely by the baryonic EOS. 

Although the DM contributes to $P_{\rm tot}$ and $\varepsilon_{\rm tot}$, the fixed-$k_F^\chi$ prescription makes these quantities independent of $n_B$. As a result, the DM sector enters $\Gamma$ only through the prefactor $(\varepsilon_{\rm tot}+P_{\rm tot})/P_{\rm tot}$, merely rescaling its magnitude without altering its density dependence (see Fig.~1, bottom panel of \cite{Lopes:2024ixl}). The present framework ensures that all DM contributions evolve with $n_B$, entering both the prefactor and the thermodynamic derivatives. The resulting modification of both the magnitude and functional form of $\Gamma$ demonstrates that the thermodynamic response is fundamentally altered, rather than simply rescaled, by the coupled BM--DM system.

\section{Results}\label{result}

\begin{table*}[t!]
\scriptsize
\centering
\caption{Structural properties of DANSs predicted by the NL3$\omega\rho$, DD2, and FSU2R RMF models for different $F_\chi$. Here, $M_{\rm max}$ and $M_{B,\rm max}$ denote the maximum gravitational and baryonic masses, respectively, while $\Delta M_{\rm max}$ is the corresponding percentage reduction in the maximum gravitational mass relative to the DM-free ($F_\chi=0$) configuration for each EOS. The quantities $R$, $\varepsilon_c$, $n_{c,\rm max}$, and $M_0$ are the radius, central energy density, central baryon density of the maximum-mass configuration, and the total rest mass, respectively. The last two columns give the radius, $R_{1.4}$, and central baryon density, $n_{c,1.4}$, of a canonical $1.4\,M_\odot$ NS.}
\begin{ruledtabular}
\begin{tabular}{ccccccccccc}
\hline
Model &
$F_\chi$ (\%) &$M_{\rm max}$ ($M_\odot$) &$\Delta M_{\rm max}$ (\%) &$M_{B,\rm max}$ ($M_\odot$) &$R$ (km) &$\varepsilon_c$ (MeV fm$^{-3}$) &$n_{c,\rm max}$ (fm$^{-3}$) &$R_{1.4}$ (km) & $n_{c,1.4}$ (fm$^{-3}$)& $M_0$($M_\odot$)\\
\hline

\multicolumn{11}{c}{\textbf{NL3$\omega\rho$}}\\
\hline

NL3$\omega\rho$
& 0.00 & 2.76 & 0.00  & 3.40 & 13.08 & 889.95  & 0.69 & 14.05 & 0.29 &3.40 \\
& 0.05 & 2.55 & 7.67  & 2.82 & 12.19 & 1022.16 & 0.72 & 13.08 & 0.31& 3.12 \\
& 0.10 & 2.37 & 14.20 & 2.39 & 11.44 & 1160.29 & 0.75 & 12.34 & 0.33& 2.90 \\
& 0.20 & 2.08 & 24.76 & 1.75 & 10.21 & 1466.14 & 0.81 & 11.12 & 0.38 & 2.50\\

\hline

\multicolumn{10}{c}{\textbf{DD2}}\\
\hline

DD2
& 0.00 & 2.45 & 0.00  & 2.97 & 12.00 & 1067.54 & 0.83 & 13.17 & 0.34 & 2.97 \\
& 0.05 & 2.25 & 7.94  & 2.47 & 11.14 & 1239.22 & 0.88 & 12.30 & 0.37 & 2.73\\
& 0.10 & 2.09 & 14.67 & 2.08 & 10.41 & 1424.71 & 0.92 & 11.54 & 0.41 & 2.52 \\
& 0.20 & 1.82 & 25.47 & 1.54 & 9.28 & 1815.41 & 1.01 & 10.36 & 0.50 & 2.20 \\

\hline

\multicolumn{10}{c}{\textbf{FSU2R}}\\
\hline

FSU2R
& 0.00 & 2.05 & 0.00  & 2.40 & 11.67 & 1152.39 & 0.94 & 12.77 & 0.38 & 2.40\\
& 0.05 & 1.86 & 9.18  & 1.96 & 10.77 & 1343.82 & 1.00 & 11.90 & 0.44 & 2.17\\
& 0.10 & 1.71 & 16.93 & 1.63 & 10.03 & 1539.15 & 1.05 & 11.04 & 0.51 & 1.98\\
& 0.20 & 1.45 & 29.27 & 1.18 & 8.99 & 1967.35 & 1.16 & 9.50 & 0.80 & 1.68\\

\hline
\end{tabular}
\end{ruledtabular}
\label{tab:maxmass}
\end{table*}
\Cref{tab:maxmass} summarizes the structural properties of DANSs predicted by the NL3$\omega\rho$, DD2, and FSU2R EOSs for different $F_\chi$. Across all models, increasing $F_\chi$ systematically reduces the $M_{\rm max}$ and $M_{\rm B,\, max}$ while decreasing the $R$ of both the  $M_{\rm max}$ and canonical $1.4\,M_\odot$ NSs, accompanied by monotonic increases in the $\varepsilon_c$, and $n_{c,\rm max}$. These correlated changes indicate that the DM admixture softens the EOS, driving NSs toward more compact configurations that require higher central densities to maintain hydrostatic equilibrium. From $F_\chi=0$ to $0.20\%$, $M_{\rm max}$ decreases by approximately $25$--$29\%$, with the largest reduction occurring for the softer FSU2R EOS (29.3\%) and the smallest for DD2 (25.5\%), whereas the intrinsically stiffer NL3$\omega\rho$ EOS remains the least affected. The $R$ of the $M_{\rm max}$ configuration contracts by about $2$--$3$ km, while $\varepsilon_c$ and $n_{c,\rm max}$ increase by up to $\sim70\%$ and $\sim20\%$, respectively. 

Similar behavior is found for canonical $1.4\,M_\odot$ NSs, demonstrating that the DM effect extends throughout the stable stellar sequence. These trends arise from the density-dependent BM--DM coupling mediated by the $Z^\prime$ vector field, whose self-consistent evolution with $n_B$ and $n_\chi=F_\chi n_B$ increases the energy cost of compression, shifting stellar equilibrium toward denser and more compact configurations. As shown in \cref{MR_curve}, moderate $F_\chi$ remain compatible with current mass--radius constraints, whereas larger values of $F_\chi$ produce configurations that fail to satisfy the observed $\sim2\,M_\odot$ lower limit. Consequently, massive pulsars provide a stringent upper bound on the amount of DM that can be accumulated inside NSs, highlighting the potential of multimessenger observations to constrain their DM content.

Although $F_\chi$ is defined as a particle-number ratio and is therefore numerically small ($0.05$--$0.2\%$), the corresponding DM particle mass fraction is amplified by the mass hierarchy, $m_\chi/m_N \approx 213$, so $F_\chi m_\chi/m_N$ reaches several tens of percent at $F_\chi=0.2\%$. Thus, the macroscopic evolution of the star is controlled by the DM contribution to the total energy density rather than by its particle abundance. Because the WIMPs are non-relativistic over the density range considered, they contribute predominantly through their rest-mass energy while providing comparatively little pressure support. This provides the microscopic origin of the pronounced EOS softening and the $\sim29\%$ reduction in the maximum NS mass predicted by the present model.

\subsection{Microphysics}\label{mcphy}
\begin{figure}[t!]
   \centering	
	\includegraphics[width=0.48\textwidth]{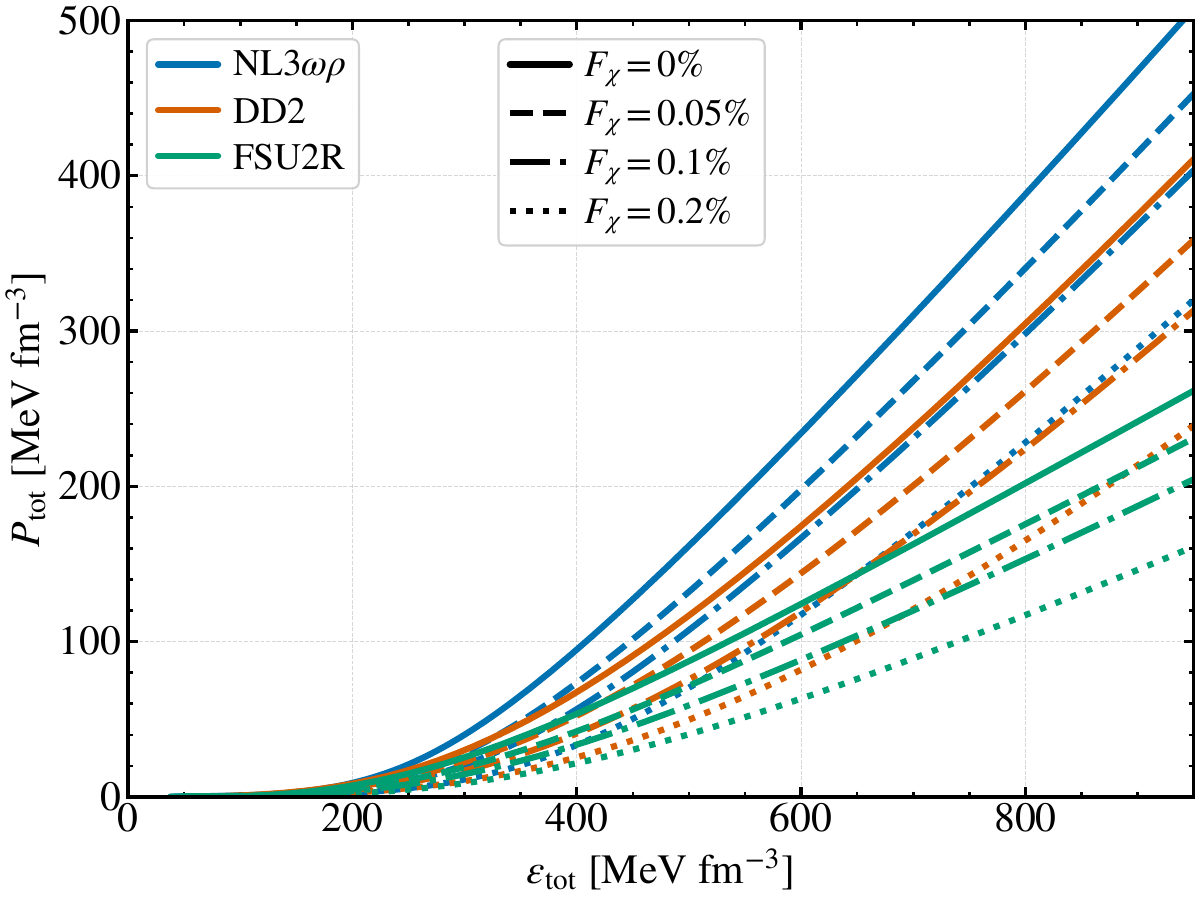}
    	\caption{Pressure as a function of the total energy density for DM-admixed stellar matter described by the NL3$\omega\rho$, DD2, and FSU2R EOSs within the self-consistent Higgs portal model. Colors distinguish the RMF parametrizations, while line styles correspond to different $F_\chi$. The systematic evolution of the curves with increasing $F_\chi$ demonstrates that the coupled Higgs and dark vector fields modify the thermodynamic response of dense matter, leading to a progressive softening of the EOS. Among the models considered, NL3$\omega\rho$ predicts the stiffest EOS, FSU2R the softest, and DD2 exhibits an intermediate behavior.
        } 
\label{EOS_curve}
\end{figure}
\Cref{EOS_curve} shows the $P_{\rm tot}$ as a function of the $\varepsilon_{\rm tot}$ for the NL3$\omega\rho$, DD2, and FSU2R EOSs. Increasing $F_\chi$ systematically shifts the EOS toward lower $P_{\rm tot}$ at a fixed $\varepsilon_{\rm tot}$, indicating a progressive softening of dense matter that becomes more pronounced at high densities. {This behavior arises because increasing $F_\chi$ self-consistently raises the $n_\chi$, thereby increasing the contribution of heavy DM particles to the total rest-mass energy, $\varepsilon_0 = n_B m_N + n_\chi m_\chi$. Since the DM sector contributes much more strongly to the energy density than to the pressure, due to the large mass hierarchy $m_\chi \gg m_N$, the $\varepsilon_{\rm tot}$ increases more rapidly than the $P_{\rm tot}$, leading to a progressive softening of the EOS.} A similar softening has been reported in other DM-admixed models~\cite{Sen:2021wev, Baym:2018ljz, Narain:2006kx, Das:2020vng}, where the additional DM component increases the energy content more rapidly than the pressure, thereby reducing the pressure support against gravity. Although the magnitude of the effect depends on the underlying RMF parameterization, the ordering of the EOSs remains unchanged, with NL3$\omega\rho$ predicting the stiffest EOS, FSU2R the softest, and DD2 an intermediate behavior. The common evolution across all three models demonstrates that the DM-induced softening is a robust feature of DM-admixed matter rather than a consequence of a particular nuclear interaction.

\begin{figure}[t!]
   \centering	
	\includegraphics[width=0.48\textwidth]{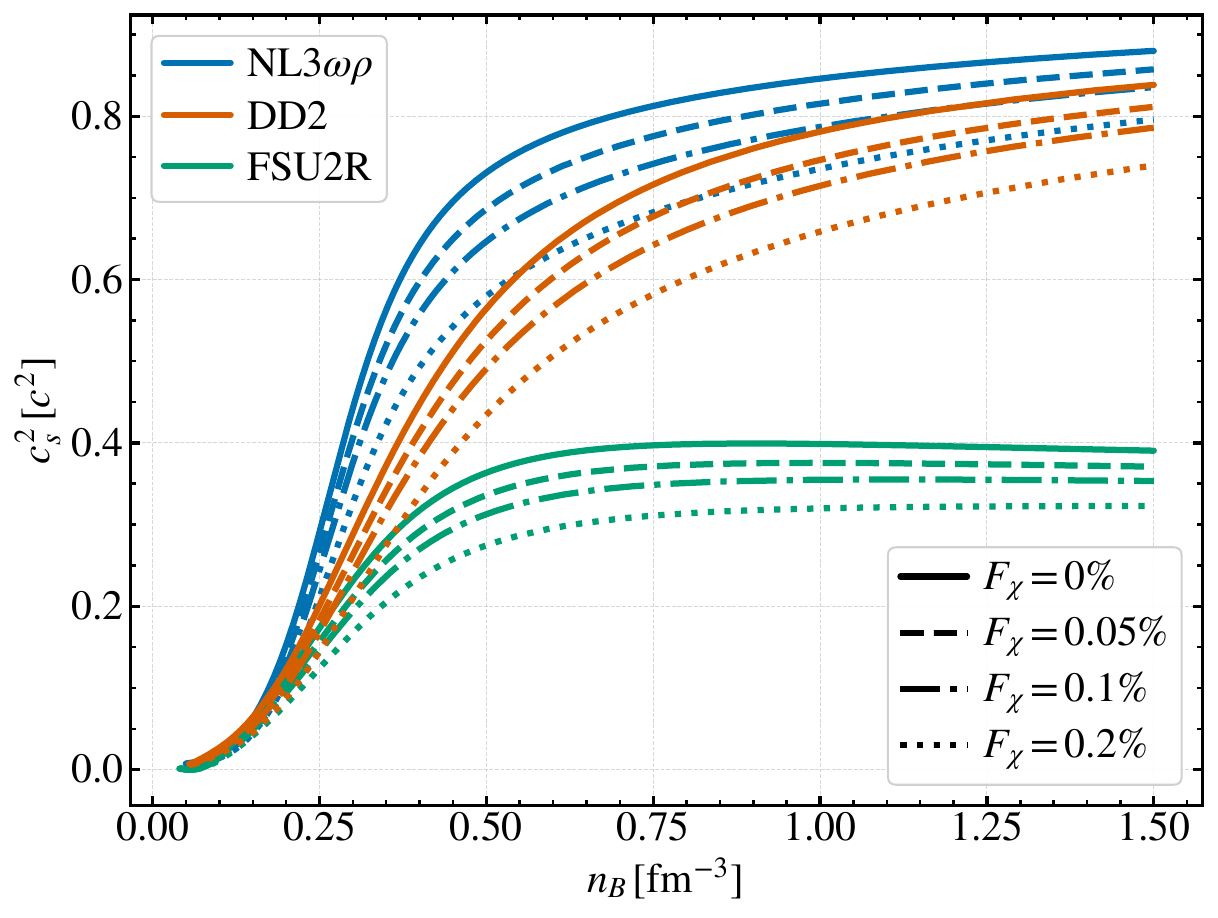}
    	\caption{The figure shows the variation of $c_s^2$ as a function of $n_B$. The results demonstrate that the presence of DM systematically reduces $c_s^2$ in dense matter across all RMF parameterizations considered, indicating a progressive softening of the EOS. The colors distinguish the different RMF parameterizations, while the line styles correspond to different $F_\chi$.} 
\label{cs2_curve}
\end{figure}
\Cref{cs2_curve} shows the $c_s^2$ as a function of $n_B$ for the three RMF EOSs. Increasing $F_\chi$ systematically suppresses $c_s^2$, with the largest deviations occurring at high densities where the BM--DM coupling is strongest, reflecting the progressive softening of the EOS. The reduction is most pronounced for the softer FSU2R EOS and weakest for the stiffer NL3$\omega\rho$ EOS. Unlike the single-fluid approaches that prescribe a fixed $k_F^\chi$, the present framework allows the DM sector to contribute dynamically to the thermodynamic derivatives, making the suppression of $c_s^2$ a direct consequence of the coupled BM--DM dynamics. Since $c_s^2$ governs the propagation of pressure perturbations and plays a central role in determining the tidal response of NS matter, its systematic reduction provides a microscopic signature of DM accumulation with direct implications for gravitational-wave observations \cite{Koehn:2024gal}.

\begin{figure}[t!]
   \centering	
	\includegraphics[width=0.48\textwidth]{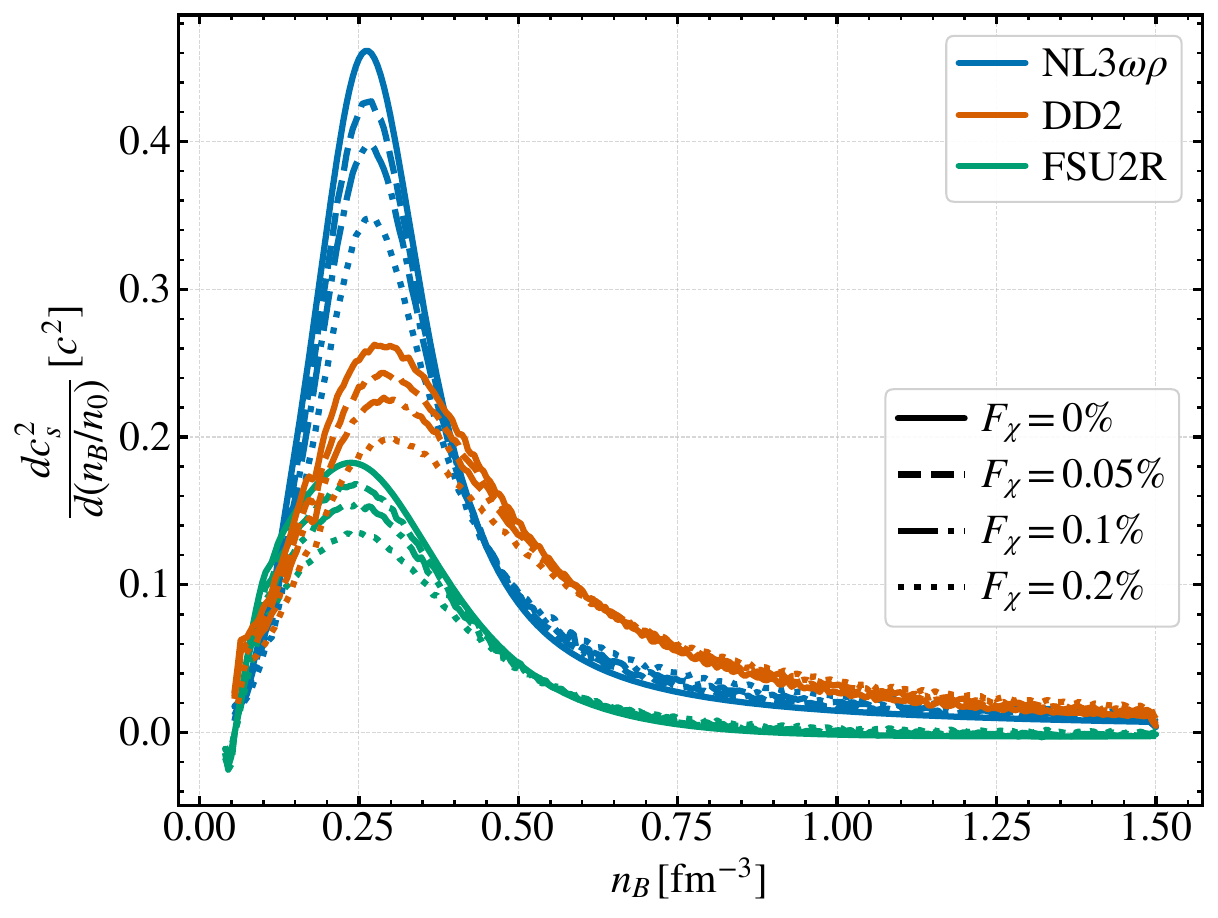}
    	\caption{Derivative of the squared speed of sound with respect to the normalized baryon density, $n_B/n_0$, for the NL3$\omega\rho$, DD2, and FSU2R EOSs. Colors distinguish the RMF parameterizations, while line styles correspond to different $F_\chi$. The systematic evolution of the curves as $F_\chi$ increases demonstrates that the DM sector modifies the density dependence of the EOS, providing an additional thermodynamic signature of DANSs.} 
\label{dcs2_curve}
\end{figure}

\Cref{dcs2_curve} shows the density derivative of $c_s^2$ for the three RMF EOSs. Increasing $F_\chi$ systematically suppresses $dc_s^2/d(n_B/n_0)$, particularly at higher densities, reflecting the progressive softening of dense matter EOS induced by the BM--DM coupling. The characteristic peak at intermediate densities decreases monotonically with increasing $F_\chi$, indicating that the thermodynamic response evolves more gradually as the DM content increases. As shown in \cref{dcs}, this behavior originates from the additional contributions to $P''$ and $\varepsilon''$ arising from the evolution of $n_\chi$ and $k_F^\chi$. Unlike single-fluid models with a fixed $k_F^\chi$, the present framework captures the dynamical contribution of the DM sector to the thermodynamic derivatives, making $dc_s^2/d(n_B/n_0)$ a sensitive microscopic probe of DM accumulation. Although the influence of DM on the EOS has been extensively investigated~\cite{Grippa:2024ach, Shirke:2025lsu}, explicit analyses of the density evolution of $c_s^2$ in DANSs remain scarce. Since the $c_s^2$ governs the propagation of pressure perturbations and plays a central role in determining the stellar structure and tidal response of NSs~\cite{Tews:2019cap, Mroczek:2023zxo}, its density derivative provides an additional thermodynamic observable for constraining the BM--DM interaction through future gravitational-wave and X-ray observations.

\subsection{Macrophysics}\label{macphy}
\begin{figure}[t!]
   \centering	
	\includegraphics[width=0.48\textwidth]{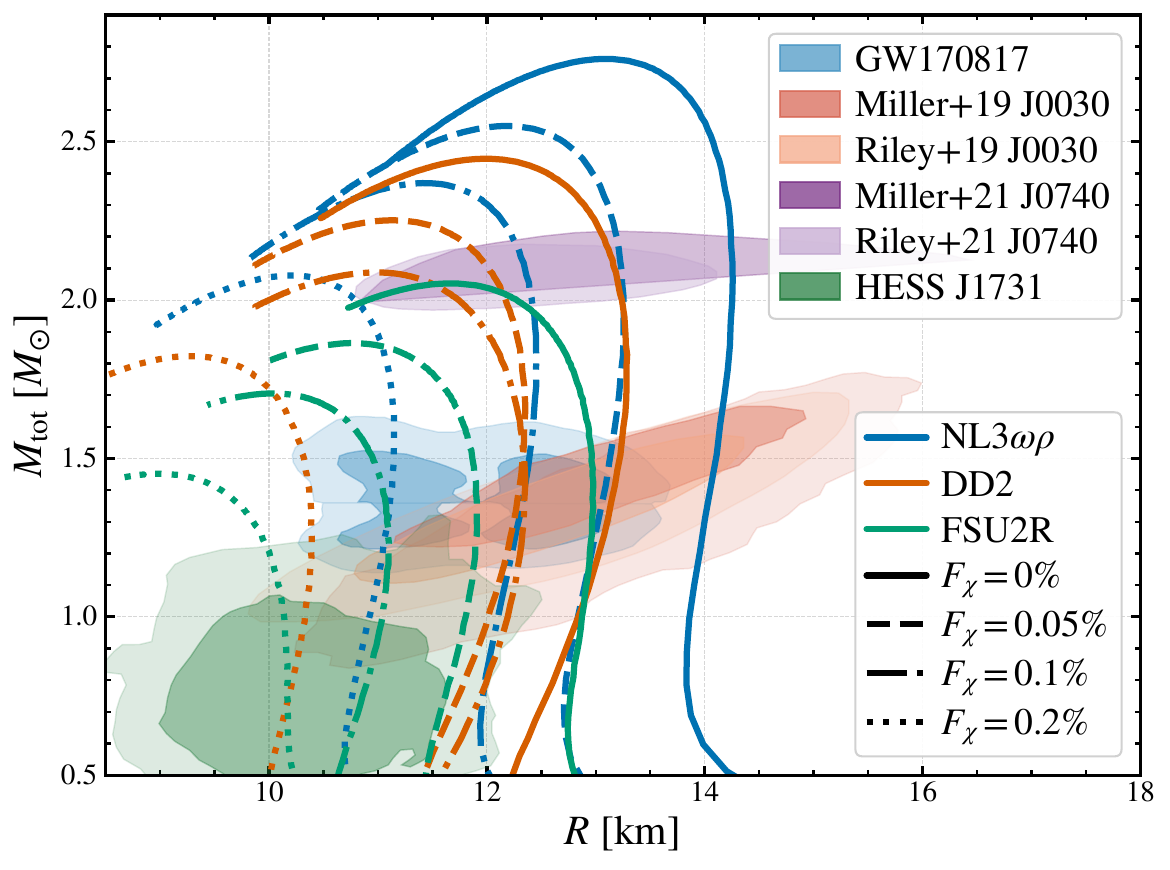}
    	\caption{The mass--radius relation predicted by the NL3$\omega\rho$, DD2, and FSU2R EOSs for different DM fractions, $F_\chi$. Colors distinguish the RMF parameterizations, while line styles correspond to the adopted DM fractions. Increasing $F_\chi$ leads to more compact stellar configurations, reflecting the modification of the equation of state induced by the coupled Higgs and dark vector interactions. Confidence contours show observational constraints from the secondary component of the binary merger event GW170817~\cite{abbott2019PhRvX} (blue; outer contour 90\% credible region (CR), inner contour 50\% CR), HESS~J1731--347~\cite{doroshenko2022strangely} (green; outer contour 95\% CR, inner contour 68\% CR), PSR~J0740+6620 (purple for Miller \textit{et al.}~\cite{miller2021} and lavender for Riley \textit{et al.}~\cite{riley2021}, both at 95\% CR), and PSR~J0030+0451 (brick for Miller \textit{et al.}~\cite{Miller:2019cac} and salmon for Riley \textit{et al.}~\cite{riley2019}, both at 95\% CR).} 
\label{MR_curve}
\end{figure}
\Cref{MR_curve} shows the mass--radius relations predicted by the three RMF EOSs for different $F_\chi$. Increasing $F_\chi$ systematically shifts the stellar sequences toward lower masses and smaller radii, consistent with the progressive softening of the EOS induced by the BM--DM coupling. The effect is strongest for the softer FSU2R EOS and weakest for the stiffer NL3$\omega\rho$ EOS, demonstrating that DM-induced softening is amplified in softer EOS. Comparison with multimessenger constraints \cite{miller2021, riley2021, Miller:2019cac, riley2019, abbott2019PhRvX} shows that moderate $F_\chi$ values remain broadly compatible with current observations (particularly DD2 and NL3$\omega\rho$ models), whereas larger values of $F_\chi$ produce configurations that become either too compact or fail to satisfy the observed $\sim2\,M_\odot$ threshold. The systematic reduction in stellar masses also brings part of the stable sequence closer to the HESS~J1731--347 CR, suggesting that moderate WIMP DM admixtures may provide a viable explanation for low-mass compact objects \cite{doroshenko2022strangely}. Overall, these results show that multimessenger observations can simultaneously constrain the nuclear EOS and the amount of DM that can be accumulated inside NSs, with softer EOSs placing the most stringent limits on $F_\chi$. %\cite{Rezaei:2016zje, Shakeri:2022dwg, Kain:2021hpk, Das:2020vng, Gresham:2018rqo}.

\begin{figure}
    \centering
    \includegraphics[width=0.48\textwidth]{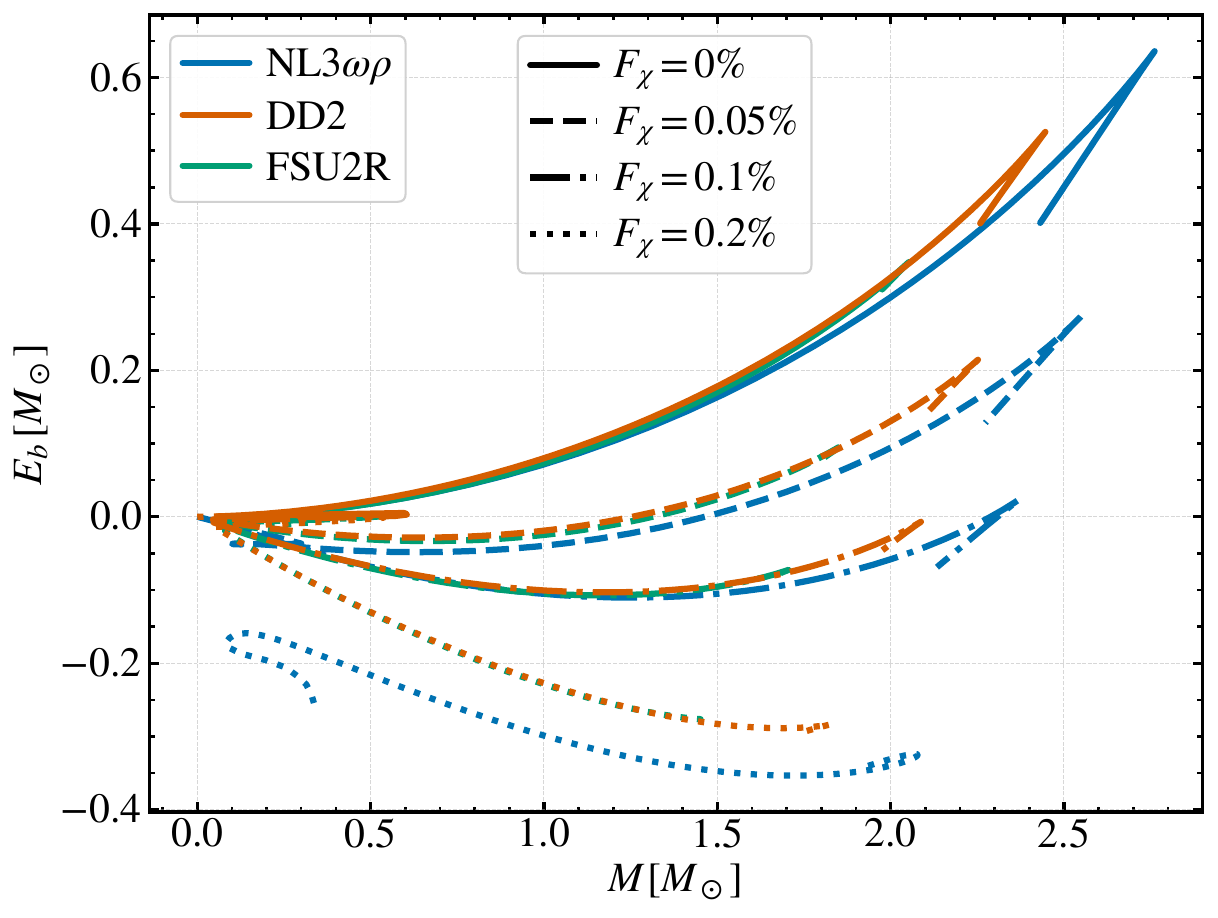}
    \caption{Gravitational binding energy, $E_b=M_B-M$, as a function of the gravitational mass for the NL3$\omega\rho$, DD2, and FSU2R EOSs at different $F_\chi$. Colors distinguish the RMF parameterizations, while line styles denote the $F_\chi$. The systematic evolution of the curves with increasing $F_\chi$ demonstrates that DM accumulation modifies $E_b$ of NSs through the BM--DM coupling, with the magnitude of the effect depending on the underlying nuclear EOS.}
    \label{fig:bind}  
\end{figure}
Figure~\ref{fig:bind} shows the gravitational binding energy, $E_b=M_B-M$, as a function of $M$ for the three RMF EOSs at different $F_\chi$. For the DM-free sequences, $E_b$ increases monotonically with $M$, consistent with the stronger gravitational confinement of more massive NSs \cite{Lattimer:2000nx}. Increasing $F_\chi$ systematically shifts the sequences toward lower masses and smaller binding energies, with $E_b$ becoming negative over part of the stable branch for $F_\chi=0.2\%$. Unlike ordinary NSs, where $E_b$ remains positive, this behavior arises from the different evolution of $M_B$ and $M$ (see \cref{tab:maxmass}): although both decrease with increasing $F_\chi$, $M_B$ decreases more rapidly, causing the ratio $M_B/M$ to fall below unity at higher values of $F_\chi$. 

Physically, the heavy WIMP component contributes significantly to $M$ while contributing nothing to the visible-sector baryon number, so $M_B$ no longer tracks the total rest-mass energy, $\varepsilon_0 = n_Bm_N + m_\chi n_\chi=n_B(m_N+F_\chi m_\chi)$, of the star. Consequently, the negative values of $E_b$ do not imply a loss of hydrostatic stability, which is still determined by the standard turning-point criterion, $dM/d\varepsilon_c>0$ \cite{Shapiro:1983du}, but instead signal the emergence of a DM-dominated contribution to the stellar mass budget. The largest absolute variation in $E_b$ is found for the NL3$\omega\rho$ EOS, despite exhibiting the smallest relative reduction in $M_{\rm max}$, whereas the DD2 and FSU2R predictions remain remarkably similar. These results show that the $E_b$ is governed not only by the EOS but also by the self-consistent evolution of the coupled BM--DM system.

We emphasize that the negative values of $E_b$ are a consequence of the adopted definition rather than an indication that the stellar configuration is gravitationally unbound. The conventional definition of $E_b$, is appropriate for ordinary NSs because the rest-mass budget is entirely determined by the baryon mass, $M_B=N_Bm_N$. In the present case, however, the star also contains a population of heavy WIMPs, so the total rest mass, $M_0$, of all constituent particles satisfies \cref{mrest}. For the benchmark WIMP mass adopted here, using $F_\chi=0.2\%$, one obtains $F_\chi(m_\chi/m_N)\simeq0.426$, implying that the DM increases the $M_0$ budget by approximately $42.6\%$. Consequently, the physically relevant $E_b$ should be defined relative to the $M_0$ of all constituent particles, $E_b^{\rm true}=M_0-M,$ which remains positive throughout the stable stellar sequence (see \cref{tab:maxmass}). Therefore, the negative values of $E_b$ simply reflect that the conventional $M_B$ definition neglects the substantial rest-mass contribution of the DM component. They do not imply a loss of hydrostatic equilibrium or gravitational binding, which continue to be determined by the standard stability criterion, $dM/d\varepsilon_c>0$.

\begin{figure}[t!]
   \centering	
	\includegraphics[width=0.48\textwidth]{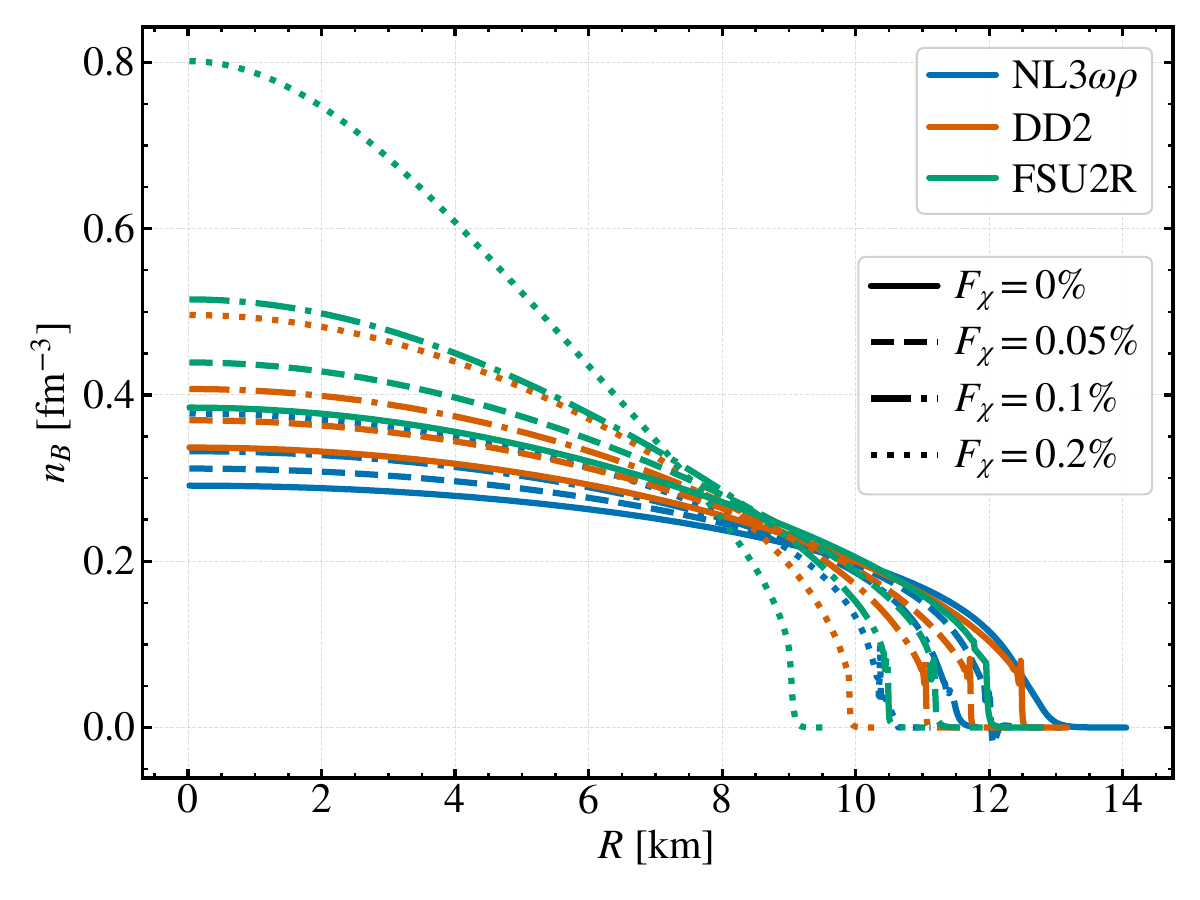}
    	\caption{The figure shows the baryon density profile as a function of the stellar radius for a canonical $1.4\, M_\odot$ NS. Increasing $F_\chi$ leads to progressively more compact stellar configurations, resulting in smaller radii and higher central baryon densities, a consequence of the DM-induced modification of the EOS. Colors distinguish the RMF parameterizations, while line styles correspond to the different $F_\chi$.} 
\label{nxr_curve}
\end{figure}
\Cref{nxr_curve} shows the baryon density profiles \cite{Glendenning2000} of canonical $1.4\,M_\odot$ DANSs for the three RMF EOSs at different $F_\chi$. Increasing $F_\chi$ systematically shifts the profiles toward smaller radii and higher central densities, reflecting the progressive compaction of the star induced by the BM--DM coupling. For the stiff NL3$\omega\rho$ EOS, $R_{1.4}$ decreases from $14.05$ to $11.12$ km, while $n_{c,1.4}$ increases from $0.29$ to $0.38~\mathrm{fm}^{-3}$ as $F_\chi$ increases from $0$ to $0.2\%$. In contrast, the softer FSU2R EOS exhibits a much stronger response, with $R_{1.4}$ shrinking from $12.77$ to $9.50$ km and $n_{c,1.4}$ increasing from $0.38$ to $0.80~\mathrm{fm}^{-3}$, indicating that DM-induced compaction is amplified in softer EOS. Since $n_\chi=F_\chi n_B(r)$, the DM distribution naturally follows the baryonic profile and becomes increasingly concentrated toward the stellar core, reproducing the expected outcome of DM capture, thermalization, and gravitational settling. This constitutes a key improvement over fixed-$k_F^\chi$ prescriptions, where the $n_\chi$ profile is prescribed independently of the local baryonic environment rather than emerging self-consistently from the coupled BM--DM field equations.

\subsection{DM--MB interaction properties}\label{dmbmint}

\begin{figure}[t!]
   \centering	
	\includegraphics[width=0.48\textwidth]{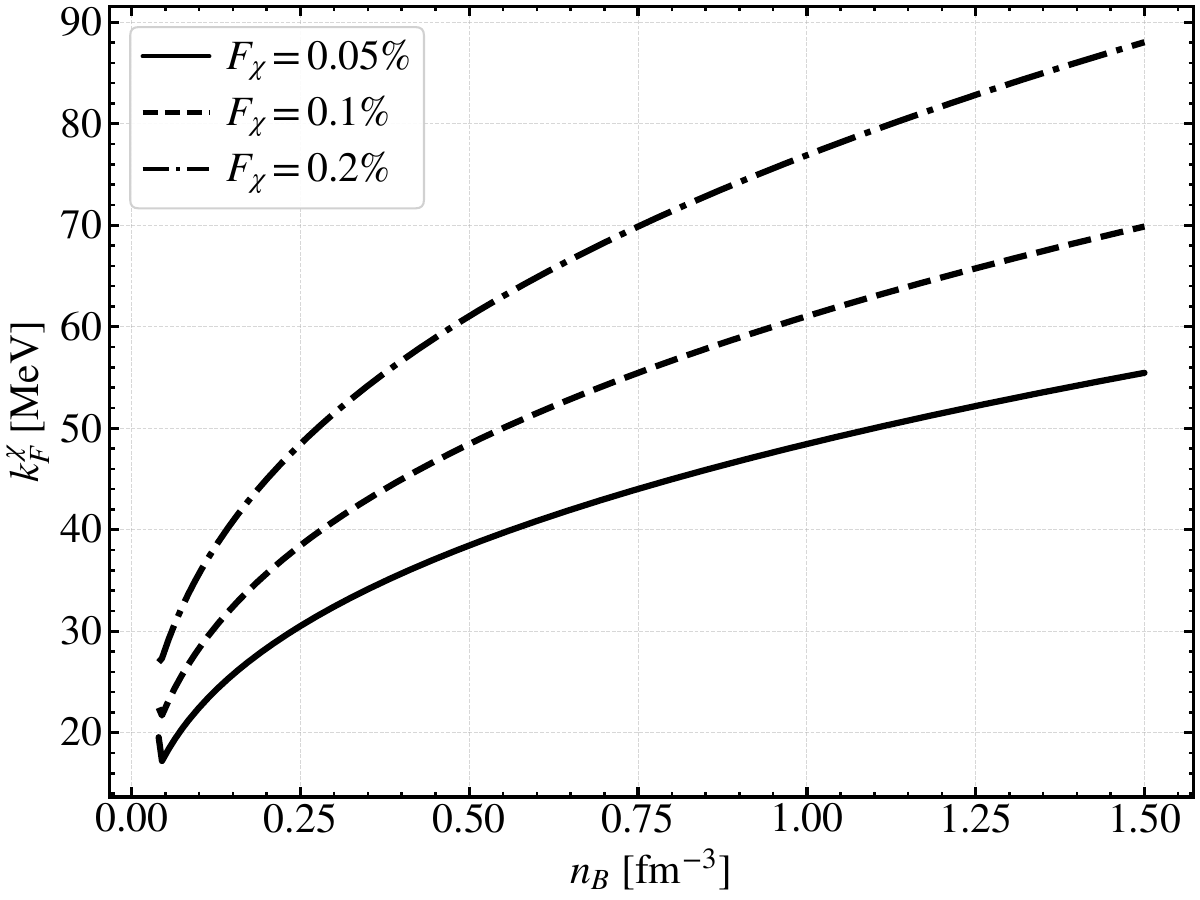}
    	\caption{The figure shows the variation of $k_F^\chi$, as a function of $n_B$ for the NL3$\omega\rho$, DD2, and FSU2R EOSs. Colors distinguish the RMF parameterizations, while line styles correspond to different  $F_\chi$. The monotonic increase of $k_F^\chi$ with both $n_B$ and $F_\chi$ reflects the self-regulated accumulation of DM in the stellar core. Since $n_\chi=F_\chi n_B$, the overall evolution of $k_F^\chi$ is largely model independent, while the underlying EOS governs its detailed density distribution.} 
\label{kfxnB_curve}
\end{figure}
\Cref{kfxnB_curve} shows $k_F^\chi$ as a function of $n_B$ for different DM fractions, $F_\chi$. Unlike single-fluid models, where $k_F^\chi$ is prescribed as an external parameter, the present framework determines $k_F^\chi$ self-consistently through $k_F^\chi=\left(3\pi^2F_\chi n_B\right)^{1/3}$,
% \begin{equation}
% k_F^\chi=\left(3\pi^2F_\chi n_B\right)^{1/3},
% \end{equation}
such that it increases monotonically with both $n_B$ and $F_\chi$, reflecting the progressive accumulation of DM toward the stellar core. Since this relation contains no explicit dependence on the RMF EOS, the predictions for NL3$\omega\rho$, DD2, and FSU2R are analytically identical, with the EOS influencing only the density profile through $n_B(r)$. The predicted values, $k_F^\chi\lesssim100$ MeV, show that the DM remains deeply nonrelativistic for the adopted WIMP mass, $m_\chi=200$ GeV, validating the nonrelativistic treatment of the DM sector. Consequently, $k_F^\chi$ emerges as a density-dependent quantity that naturally traces the local $n_B$ environment, providing a physically motivated description of DM capture, thermalization, and gravitational settling, in contrast to fixed-$k_F^\chi$ prescriptions.

\begin{figure}[t!]
   \centering	
	\includegraphics[width=0.48\textwidth]{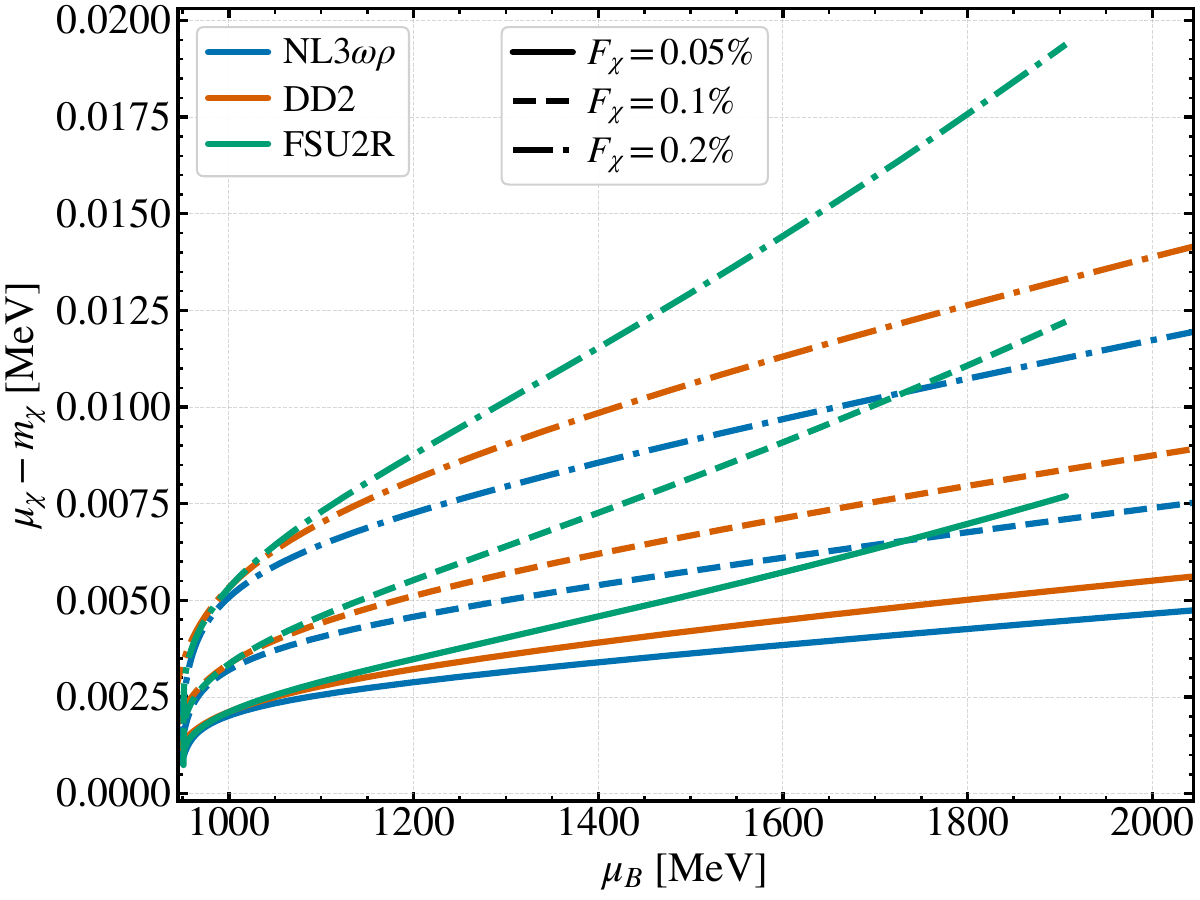}
    	\caption{The figure shows the variation of $\mu_\chi-m_\chi$ as a function of the baryon chemical potential, $\mu_B$, for the NL3$\omega\rho$, DD2, and FSU2R EOSs. Colors distinguish the RMF parameterizations, while line styles correspond to different $F_\chi$. The quantity $\mu_\chi-m_\chi$ increases monotonically with $\mu_B$, reflecting the increase in the DM kinetic contribution to the chemical potential as the stellar density rises. However, the overall variation remains modest, of the order of $10^{-3}-10^{-2}$ MeV, indicating that the DM chemical potential remains close to the DM rest mass over the density range considered.} 
\label{muDxmuB_curve}
\end{figure}
\begin{figure*}[t!]
   \centering	
	\includegraphics[width=1\textwidth]{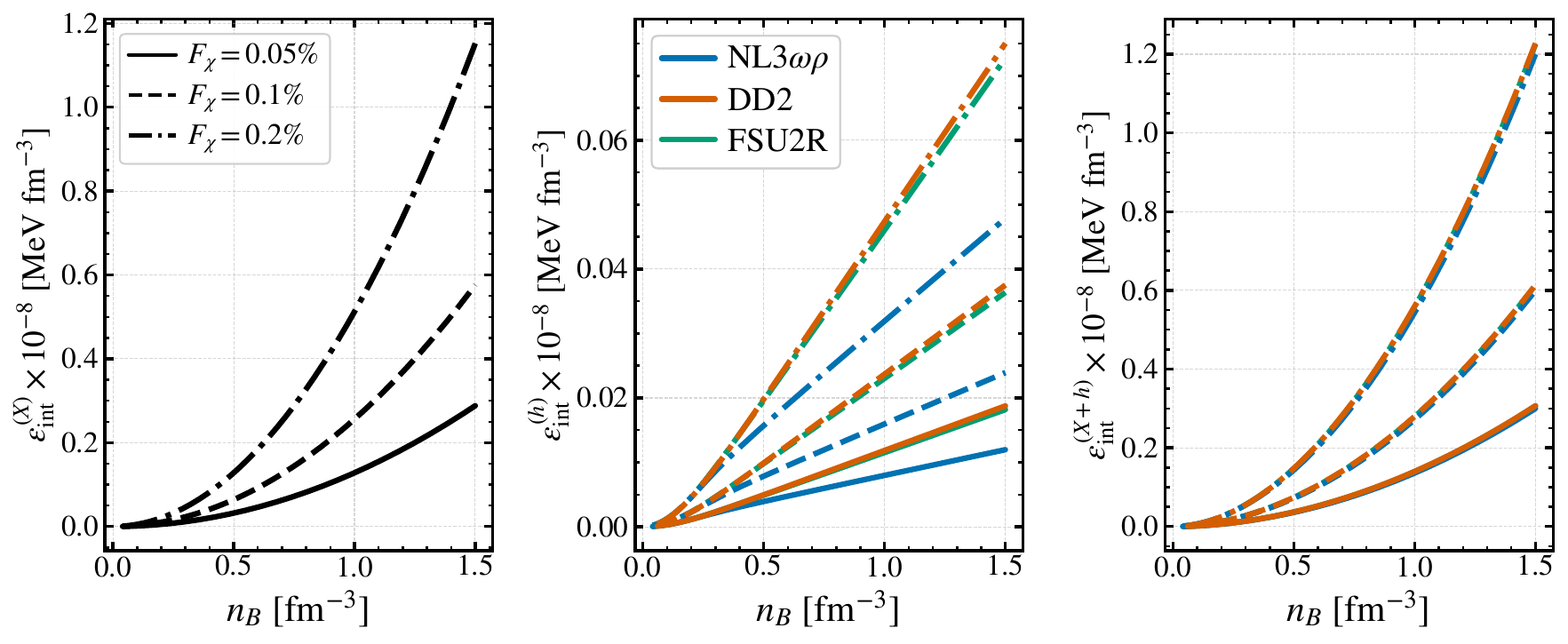}
    	\caption{The figure shows the vector, scalar (Higgs), and total interaction energy densities as functions of $n_B$ for the NL3$\omega\rho$, DD2, and FSU2R EOSs. The left, middle, and right panels correspond to the vector interaction energy density, $\varepsilon_{\rm int}^{(X)}$, the scalar interaction energy density, $\varepsilon_{\rm int}^{(h)}$, and their total contribution, $\varepsilon_{\rm int}^{(X+h)}$, respectively. Colors distinguish the RMF parameterizations, while line styles correspond to different $F_\chi$. The $\varepsilon_{\rm int}^{(X)}$ is nearly model-independent, being determined primarily by the local $n_B$ and $n_\chi = F_\chi n_B$, whereas the $\varepsilon_{\rm int}^{(h)}$ exhibits a stronger EOS dependence through the Higgs-induced modification of the scalar densities. Consequently, the $\varepsilon_{\rm int}^{(X+h)}$ inherits the EOS dependence of the scalar sector while increasing systematically with both $n_B$ and $F_\chi$. } 
\label{int}
\end{figure*}

\Cref{muDxmuB_curve} shows the variation of $\mu_\chi-m_\chi$ as a function of $\mu_B$ for the three RMF EOSs at different $F_\chi$. In the nonrelativistic limit \cref{ChemPot} expands as,
\begin{equation}
\mu_\chi-m_\chi\simeq-g_{h}h_0+g_{\chi X}X_0+\frac{\left(3\pi^2F_\chi n_B\right)^{2/3}}{2m_\chi^*},
\label{eq:muchi_shift}
\end{equation}
showing that the chemical-potential shift receives contributions from the Higgs-induced effective mass, the repulsive $Z^\prime$ vector mean-field, and the DM Fermi kinetic energy. Since the kinetic term scales as $n_B^{2/3}/m_\chi^*$ whereas the vector contribution grows approximately linearly with $n_B$ (see \cref{Xfield}), the monotonic increase of $\mu_\chi-m_\chi$ with $\mu_B$ is primarily driven by the density-dependent vector interaction. Nevertheless, its magnitude remains only $10^{-3}$--$10^{-2}$ MeV, more than ten orders of magnitude smaller than the adopted WIMP mass, $m_\chi=200$ GeV. Together with $k_F^\chi\lesssim90$ MeV (\cref{kfxnB_curve}), this indicates that the DM remains deeply nonrelativistic, with $\mu_\chi\simeq m_\chi$ throughout the stellar interior. Consequently, the DM contributes predominantly through its rest-mass energy, while its Fermi pressure remains comparatively small. The modest spread among the RMF EOSs originates solely from their different $n_B(\mu_B)$ relations, since the dark-sector parameters are fixed independently of the nuclear model. These microscopic results explain the EOS softening (see \cref{EOS_curve,cs2_curve}), the dominance of the vector interaction energy (see \cref{int}), and the emergence of negative $E_b$ at large $F_\chi$ (\cref{fig:bind}), demonstrating that the macroscopic properties of DANSs are governed primarily by the heavy WIMP rest-mass contribution and the BM--DM interaction rather than by DM degeneracy pressure.

\Cref{int} presents the vector (left), scalar (middle), and total (right) BM--DM interaction energy densities as functions of $n_B$ for the three RMF EOSs at different $F_\chi$. The vector interaction energy, $\varepsilon_{\rm int}^{(X)}$, increases systematically with $n_B$ and $F_\chi$ and is nearly EOS independent because $\varepsilon_{\rm int}^{(X)}\propto F_\chi n_B^2$, which depends explicitly only on the local $n_B$ through $n_\chi=F_\chi n_B$. In contrast, the scalar interaction energy, $\varepsilon_{\rm int}^{(h)}\propto n_s^Bn_s^\chi$, exhibits a stronger EOS dependence through the scalar densities and the corresponding effective masses. Consequently, the total interaction energy inherits the modest EOS dependence of the scalar sector but is overwhelmingly dominated by the repulsive $Z^\prime$-mediated vector interaction, with $\varepsilon_{\rm int}^{(h)}/\varepsilon_{\rm int}^{(X)}\lesssim5\%$ throughout the density range considered.

This hierarchy reflects the ATLAS benchmark couplings adopted in this work, for which the vector interaction is structurally enhanced relative to the Higgs portal. The resulting dominance of the density-dependent $Z^\prime$ mean-field provides a microscopic explanation for the EOS softening (\cref{EOS_curve}). The suppression of $c_s^2$ (\cref{cs2_curve,dcs2_curve}), the enhanced compactness and binding of DANSs (\cref{MR_curve,fig:bind}), jointly corroborates the conclusions drawn from the $\mu_\chi-m_\chi$ analysis (\cref{muDxmuB_curve}). More importantly, this decomposition establishes a direct connection between the underlying particle-physics interaction channels and the macroscopic properties of DANSs, providing a quantitative framework for assessing the relative roles of scalar and vector portals in self-consistent single-fluid DANSs.

\section{Conclusions}\label{conc}

We have developed a self-consistent single-fluid framework for DANSs by extending the Higgs-portal model with a massive $Z^\prime$ vector mediator. Unlike single-fluid DANSs formulations, which prescribe $k_F^\chi$ externally, the present approach determines the local $n_\chi$ through the physically motivated relation $n_\chi=F_\chi n_B$, allowing $n_\chi$, $k_F^\chi$, and mean-fields to evolve consistently with the stellar environment. Beyond this conceptual advance, we derived exact analytical expressions for $c_s^2$, its density derivative, and $\Gamma$, explicitly identifying the baryonic, dark, scalar, and vector contributions to the thermodynamic response of dense matter. These analytical results establish a general benchmark for assessing the causality and thermodynamic stability of self-consistent DM-admixed EOSs and demonstrate that the DM sector modifies the thermodynamic response itself rather than simply contributing an externally prescribed energy component.

A central physical outcome of this framework is the identification of distinctive signatures of heavy WIMP admixture that emerge naturally from the self-consistent BM--DM coupling. The interaction-energy decomposition reveals quantitatively that the repulsive $Z^\prime$ vector channel overwhelmingly dominates the Higgs portal for the collider-motivated benchmark adopted here, providing the microscopic origin of the EOS softening, $c_s^2$ suppression, and enhanced stellar compactness. At the same time, $\mu_\chi$ remains extremely close to the WIMP rest mass, implying that the DM sector contributes predominantly through its rest-mass energy, $\varepsilon_{\chi,0} = m_\chi n_\chi = F_\chi n_B m_\chi$ while its Fermi pressure remains negligible. This hierarchy gives rise to characteristic macroscopic signatures, including the evolution of the $c_s^2$ profile and the prediction of negative gravitational binding energies at large $F_\chi$, arising from the decoupling of the $M_B$ and $M$ budgets rather than from any loss of hydrostatic stability. Together, these analytical and numerical results identify a set of microscopic and macroscopic observables that can distinguish heavy-WIMP effects from uncertainties associated with the nuclear EOS.

By adopting collider-constrained benchmark masses and couplings for both the Higgs and $Z^\prime$ portals, this work establishes a direct and quantitative connection between particle physics and multimessenger astrophysics. Within the present framework, the global $F_\chi$ becomes the only astrophysical degree of freedom, allowing observational constraints from GW170817, NICER, and HESS~J1731--347 to be translated directly into limits on the amount of DM that can accumulate inside NSs for a fixed, collider-consistent particle-physics scenario. This demonstrates how compact stars can serve as complementary laboratories for testing collider-motivated DM models and provides a unified framework in which future gravitational-wave \cite{ET:2025xjr}, X-ray \cite{watts2019dense}, and terrestrial experiments can jointly constrain the properties of DM. More broadly, the present work lays the foundation for establishing genuine two-way constraints between particle physics and multimessenger astrophysics, where collider measurements inform NS modeling and astrophysical observations, in turn, constrain viable regions of the DM parameter space.

More generally, the present work demonstrates that enforcing self-consistency in the treatment of the dark sector is essential for obtaining physically meaningful single-fluid EOSs and reliable predictions for the structure and observable properties of DANSs. Throughout the density range considered, the predicted EOS remains causal ($c_s^2<1$) and thermodynamically stable, confirming the internal consistency of the framework. Although $F_\chi$ is treated here as a free astrophysical parameter, establishing its physically realistic range through explicit calculations of DM capture, thermalization, and accumulation for the adopted WIMP mass and interaction cross section constitutes an important direction for future work. The predicted modifications to the EOS, stellar compactness, maximum mass, and thermodynamic response provide concrete observables offering a promising avenue for constraining the properties of DM in compact stars.

\section*{Acknowledgement}
A.I. acknowledges financial support from the São Paulo State Research Foundation (FAPESP), Grant Nos. 2023/09545-1 and 2025/17347-0. This work is part of the project INCT-FNA (Proc. No. 464898/2014-5).

\bibliography{references}

\end{document}